\documentclass{aa}
\usepackage[breaklinks,colorlinks,linkcolor=blue,citecolor=blue,urlcolor=blue,hyperfootnotes=true]{hyperref}
\usepackage{graphicx}
\usepackage{txfonts}
\usepackage{url}
\usepackage{graphicx}

\usepackage{xcolor}
\usepackage[normalem]{ulem}

\begin{document}

\title{Newly detected open clusters in the Galactic disk using {\textit{Gaia}} EDR3}

\author{C. J. Hao\inst{1,2}, Y. Xu\inst{1,2}, Z. Y. Wu\inst{3,4},
Z. H. Lin \inst{1,2}, D. J. Liu\inst{1,2}, Y. J. Li\inst{1}}

\institute{Purple Mountain Observatory, Chinese Academy of Sciences, Nanjing 210023, PR China
\email{xuye@pmo.ac.cn}
\and
School of Astronomy and Space Science, University of Science and Technology of China, Hefei 230026, PR China
\and
National Astronomical Observatories, Chinese Academy of Sciences, 20A Datun Road, Chaoyang District, Beijing 100101, PR China
\and
School of Astronomy and Space Science, University of Chinese Academy of Sciences, Beijing 101408, PR China
}

\date{Received 12 January 2022 / Accepted 24 January 2022}
\titlerunning{704 newly detected open clusters}
\authorrunning{C. J. Hao et al.}

\abstract
{The astrometric satellite \emph{Gaia} recently released part of its 
third data set, which provides a good opportunity to hunt for more 
open clusters in the Milky Way. In this work, we conduct a blind 
search for open clusters in the Galactic disk using a sample-based
clustering search method with high spatial resolution, which is 
especially suited to finding hidden targets. In addition to confirming 
1 930 previously known open clusters and 82 known globular clusters, 
704 new stellar clusters are proposed as potential open clusters at 
Galactic latitudes of $|b| \le $ 20$^{\circ}$. For each of these new 
open clusters, we present the coordinates, detailed astrometric 
parameters, and ages, as well as the radial velocity, if available. 
Our blind search greatly increases the number of Galactic open 
clusters as objects of study and shows the incompleteness of the 
open cluster census across our Galaxy.}


\maketitle
%

\section{Introduction}

An open cluster (OC) is a group of stars that formed in the same molecular
cloud, so its member stars have approximately the same age and are 
gravitationally bound to each other. The known ages of OCs cover a wide
range, from a few million years to billions of years, making them 
potentially good tracers for studying the structure and evolution of the Milky 
Way \citep[e.g.,][] {janes1982,friel1995,buckner2014,hao2021,castro2021a,
poggio2021}. In addition, all OCs play 
an important role in constraining stellar structure and evolutionary models 
\citep[e.g.,][]{vandenberg1983,Barnes2007,Motta2017,marino2018}. However, 
recent works have shown that the current catalog of Galactic OCs is still 
incomplete and, thus, a complete catalog\ of OCs has yet to be compiled 
\citep[e.g.,][]{cantat2018,castro2018,castro2019,castro2020,liu2019}.

Accurately searching for Galactic OCs has attracted the interest of many 
astronomers \citep[e.g.,][]{dias2002, kharchenko2013}. Since the publication of
the second data release of \emph{Gaia} \citep[hereafter \emph{Gaia} DR2;][]
{Prusti2016,Brown2018}, studies to detect new OCs and accurately determine
membership probability have continued to emerge as an ongoing stream. Based on 
\emph{Gaia} DR2, the astrometric parameters and membership probabilities
of $\sim$ 1 200 known OCs presented in previous catalogs \citep[$\sim$
3 000 objects, e.g.,][]{dias2002, kharchenko2013} were recalculated by 
\cite{cantat2018}, who also discovered 60 new OCs. 
Meanwhile, a machine-learning approach was developed using \emph{Gaia} 
Data, with 23 new OCs found in the Tycho-Gaia Astrometric Solution 
dataset \citep[see][]{castro2018}. Since then, \cite{cantat2019} 
found 41 new OCs in the direction of the Perseus Arm. Soon after, 53 new 
OCs in the direction of the Galactic anti-Centre were detected by 
\cite{castro2019}. Next, by visually inspecting the diagrams of proper motions, 
\cite{sim2019} discovered 207 OCs, while \cite{liu2019} reported 2 443 stellar 
clusters, of which 76 were novel OCs of high quality. By applying machine 
learning in a blind search, \cite{castro2020} identified 582 new OCs in the 
Galactic disk. Meanwhile, \cite{he2021} found 74 new OCs and 
\cite{ferreira2021} discovered 34 new star clusters towards the Galactic bulge. 
In addition, 41 new OC candidates were reported by \cite{hunt2021}. 
All of these efforts have greatly improved the amount of Galactic OCs known 
to astronomers and have also indicated that there are still many unnoticed 
OC objects that have yet to be cataloged.

Numerous studies also have been conducted to study the astrophysical properties 
of OCs in \emph{Gaia} and, simultaneously, to use OCs to unveil the 
structure of the Milky Way. Through a membership assignment procedure, 
\cite{cantat2020a} derived member stars of 1 481 known OCs with \emph{Gaia} 
DR2 and investigated their characteristics. In addition, based on 
\emph{Gaia} DR2 astrometry, \cite{cantat2020b} compiled a broad and 
homogeneous OC catalog, in which 1 867 clusters have reliable parameters, 
and they studied the structure and evolution of the Galactic disk using these 
OCs. \cite{tarricq2021} computed the kinematic parameters of 1 382 OCs with 
the purpose of revealing the dynamical behaviours of them over time.
Meanwhile, a homogeneous sample of fundamental parameters of Galactic 
open clusters was presented by \cite{dias2021}, where they also analyzed the 
zero-point offset of \emph{Gaia} data.

The \emph{Gaia} satellite has released part of its third data set 
(\emph{Gaia} Early Data Release 3, hereafter \emph{Gaia} EDR3), which 
contains the celestial positions, parallaxes, and proper motions ($l$, $b$, 
$\varpi$, $\mu_{\alpha^{*}}$, and $\mu_{\delta}$), in three 
photometric bands ($G$, $G_{\rm BP}$ and $G_{\rm RP}$) for more than 
1.5 billion sources. The astrometric data have been updated significantly 
relative to DR2, that is, the parallax accuracies have reached 20--30~$\mu$as
\citep{gaia2020}. Moreover, precise radial velocities (\textit{RV}s) of seven 
million stars are also available. Such a vast data store allows us to 
detect more OCs and enable further studies of the Milky Way.

Based on an unsupervised clustering algorithm, namely, a density-based 
spatial clustering of applications with noise \citep[DBSCAN,][]{ester1996}, 
\cite{castro2018} developed a methodology to find spatial clusters in a 
five-dimensional (5D) parametric space and then they used an artificial 
neural network (ANN) approach to automatically distinguish real OCs 
and statistical clusters.
Subsequently, this method was adopted in \textit{Gaia} DR2 and plenty of 
new OCs was discovered by them \citep{castro2019,castro2020}. 
Most recently, \cite{castro2021b} reported the discovery of 628 new 
open star clusters within the Galactic disk again. 
Up to now, a total of $\sim$ 1 300 new open clusters were hunted by 
them, and thus, this result fully demonstrates the power of the algorithms 
they adopted.

\cite{hao2020} also adapted the powerful algorithms applied in 
\cite{castro2018} to detect new OCs (Paper I).
Using the OC sample provided in \cite{cantat2018}, Paper I firstly
performed a statistical analysis on those known OCs to confirm that 
there is no need to weigh the 5D parameters (i.e., $l$, $b$, $\varpi$, 
$\mu_{\alpha^{*}}$ and $\mu_{\delta}$) specifically used in the DBSCAN 
algorithm.
Then, Paper I focused on smaller regions, that is, rectangles of $1^{\circ}$ 
$\times$ $1^{\circ}$, giving a higher spatial resolution to detect 
hidden OCs, and normalized the 5D parameters using 
$z$-score standardization.
Afterwards, inspired by \cite{castro2018}, Paper I also used the
Gaussian kernel density estimation algorithm \citep{Lampe2011} and 
$k$-nearest neighbors algorithm \citep{Altman1992} to obtain suitable 
parameters for the DBSCAN algorithm.

The purpose of this work is to blindly search for unnoticed, especially 
hidden, OCs in the Galactic disk using \emph{Gaia} EDR3. 
Here, we amended the sample-based clustering search method used in 
Paper I to improve its sensitivity to spatial clusters. 
The OCs were independently hunted in different distance bins and 
the sizes of the regions being searched varied automatically according 
to the spatial density of the regions.
In addition, to obtain more reliable OC candidates, we propose 
the stellar clusters as potential OCs based on their kinematics 
and photometric information.

We organize this paper as follows. Sect. \ref{sect:data} describes 
the data used in this work. In Sect. \ref{sect:search}, we show how we 
amended and applied the method in detail. The results of displaying the 
newly found OCs are presented in Sect. \ref{sect:results}. Finally, 
Sect. \ref{sect:conclusions} summarizes the results obtained in this 
work.
%


\section{Data}
\label{sect:data}

The data used in this work are from the \textit{Gaia} EDR3 catalog
\citep{gaia2020}. Most of the known OCs are located in $|b|$ $\le$ 
$20^{\circ}$ \citep[$\sim$ 96\%, e.g.,][]{dias2002,kharchenko2013,
cantat2018} and, thus, the expectation of finding OCs in this region is
a maximum. Our search focused on the Galactic disk, namely, in the region
of $0^{\circ}\le l \le 360^{\circ}$ and $-20^{\circ}\le b \le 20^{\circ}$. 
Apart from missing the fraction of bright stars at $G$ < 7, the 
\textit{Gaia} survey is complete in this region \citep[see][]{gaia2020}.

In addition, we only extracted sources brighter than $G$ = 18, which is the 
same as the criterion used in \cite{cantat2018} and \cite{castro2021b}. 
At this magnitude, the median standard uncertainties in the parallaxes 
are 0.136~mas, and those of the proper motions, $\mu_{\alpha^{*}}$ and 
$\mu_{\delta}$, are 0.145 and 0.122~mas~yr$^{-1}$, respectively 
\citep{lindegren2021a}. As in \cite{liu2019}, in order to exclude 
observational artifacts due to faintness, we also rejected all sources with 
$\varpi$ < 0 mas or $|\mu_{\alpha^{*}}|$, $|\mu_{\delta}|$ > 
30~mas~yr$^{-1}$. As a result, we obtained a large data set containing 
220\,222\,696 stars.
%


\section{Methods}
\label{sect:search}
This section provides a detailed description of the methods used to 
search for OCs in the Galactic disk and our process for consistently 
confirming the new OCs. In this work, the sample-based
clustering search method used in Paper I has been further amended.


\subsection{Sample division}
\label{sample}

Multiple OCs at different distances along the line of sight are 
superimposed, making it very difficult to accurately decompose them 
and find the hidden OCs. To solve this puzzle, we suggested in Paper I 
to make improvements on samples for the purpose of finding hidden OCs 
by focusing on small spatial regions and considering the effect of distance.
Hence, in this work, we independently searched for OCs in different 
distance bins: <1~kpc, 1--2~kpc, 2--3~kpc, 3--4~kpc, 4--5~kpc, and 
>5~kpc, where each bin includes tens of million stars. 
Simultaneously, this technique can also reduce the computational 
procedure.

Another key aspect of the sample-based clustering search method is 
to decide the spatial sizes of different regions. In our previous work, 
rectangles of $1^{\circ}$ $\times$ $1^{\circ}$ were used by Paper I. 
Here, we have made an improvement on this point, where the sizes 
varied according to the spatial density of the regions. We still divided 
the regions into squares, but the sizes of the squares in different 
distance bins are in the range of [$1^{\circ}$, $10^{\circ}$], where the 
higher the density of stars, the smaller the size.
This improved measure, which adapts to stellar population densities 
automatically, is a good solution to the problem of uneven spatial 
distributions among stars when searching for star clusters. Finally, we 
yielded about 40\,000 spatial boxes in the six distance bins.

\begin{table*}
\centering
\caption{Parameters and errors of some examples of the proposed new OCs ordered by increasing $l$. The format of table is the same as in \cite{castro2020} and \cite{castro2021b}.}
\scriptsize
\label{Table:OCs}
\setlength{\tabcolsep}{1.74mm}
\renewcommand\arraystretch{1.11}
\begin{tabular}{l r@{$\pm$}l r@{$\pm$}l r@{$\pm$}l r@{$\pm$}l l r@{$\pm$}l r@{$\pm$}l r@{$\pm$}l r@{$\pm$}l l l r l r@{(}l@{)}}
\hline \\
\multicolumn{1}{c}{ID} &
\multicolumn{2}{c}{$\alpha$} &
\multicolumn{2}{c}{$\delta$} &
\multicolumn{2}{c}{$l$} &
\multicolumn{2}{c}{$b$} &
\multicolumn{1}{c}{$\theta$} &
\multicolumn{2}{c}{$\varpi$} &
\multicolumn{2}{c}{$\mu_{\alpha^{*}}$} &
\multicolumn{2}{c}{$\mu_{\delta}$} &
\multicolumn{2}{c}{$\log$(age)} &
\multicolumn{1}{c}{$A$$_{\rm G}$} &
\multicolumn{1}{c}{$z$} &
\multicolumn{2}{c}{$V_{r}$} &
\multicolumn{2}{c}{$N$}
\\
\multicolumn{1}{c}{} &
\multicolumn{2}{c}{[deg]} &
\multicolumn{2}{c}{[deg]} &
\multicolumn{2}{c}{[deg]} &
\multicolumn{2}{c}{[deg]} &
\multicolumn{1}{c}{[deg]} &
\multicolumn{2}{c}{[mas]} &
\multicolumn{2}{c}{[mas/yr]} &
\multicolumn{2}{c}{[mas/yr]} &
\multicolumn{2}{c}{[yr]} &
\multicolumn{1}{c}{[mag]} &
\multicolumn{1}{c}{} &
\multicolumn{2}{c}{[km/s]} &
\multicolumn{2}{c}{($N_{r}$)}
 \\ \\
 \hline \\
OC-0001 & 267.98 &   0.03 & -27.85 &   0.02 &   1.64 &   0.02 &  -0.63 &   0.02 &   0.03 &   0.35 &   0.01 &   0.75 &   0.07 &   0.12 &   0.06 & 7.40 & 0.17 & 3.22 & 0.028 &   -- & (--) &   17 &  0 \\ 
OC-0002 & 263.38 &   0.07 & -25.02 &   0.05 &   1.89 &   0.05 &   4.37 &   0.06 &   0.08 &   0.82 &   0.05 &   1.16 &   0.29 &  -0.23 &   0.30 & 8.00 & 0.18 & 3.72 & 0.028 & -12.26 &   (--) &   33 &  1 \\ 
OC-0003 & 270.24 &   0.03 & -24.87 &   0.03 &   5.24 &   0.04 &  -0.88 &   0.03 &   0.05 &   0.55 &   0.02 &   1.00 &   0.09 &  -0.19 &   0.16 & 8.80 & 0.20 & 2.56 & 0.028 &   -- & (--) &   16 &  0 \\ 
OC-0004 & 270.68 &   0.02 & -24.26 &   0.03 &   5.97 &   0.02 &  -0.93 &   0.03 &   0.04 &   0.76 &   0.02 &   1.78 &   0.19 &  -1.88 &   0.14 & 6.60 & 0.15 & 1.38 & 0.016 &   -- & (--) &   17 &  0 \\ 
OC-0005 & 271.80 &   0.03 & -24.73 &   0.03 &   6.06 &   0.04 &  -2.04 &   0.02 &   0.05 &   0.55 &   0.02 &   0.59 &   0.14 &  -0.79 &   0.23 & 8.10 & 0.19 & 3.46 & 0.028 &   3.36 &   (--) &   20 &  1 \\ 
OC-0006 & 271.09 &   0.06 & -24.37 &   0.05 &   6.06 &   0.05 &  -1.31 &   0.06 &   0.07 &   0.77 &   0.04 &   1.29 &   0.32 &  -2.08 &   0.23 & 6.30 & 0.14 & 1.52 & 0.025 &   -- & (--) &  253 &  0 \\ 
OC-0007 & 268.92 &   0.02 & -22.67 &   0.03 &   6.54 &   0.03 &   1.26 &   0.02 &   0.04 &   0.34 &   0.01 &  -1.32 &   0.13 &  -3.68 &   0.11 & 8.30 & 0.19 & 4.04 & 0.028 &   -- & (--) &   23 &  0 \\ 
OC-0008 & 271.47 &   0.02 & -23.81 &   0.02 &   6.72 &   0.02 &  -1.33 &   0.02 &   0.03 &   0.52 &   0.01 &   0.27 &   0.20 &  -0.70 &   0.13 & 7.50 & 0.17 & 3.22 & 0.016 &   -- & (--) &   15 &  0 \\ 
OC-0009 & 271.60 &   0.03 & -23.25 &   0.03 &   7.26 &   0.02 &  -1.16 &   0.03 &   0.04 &   0.36 &   0.01 &  -0.04 &   0.17 &  -1.98 &   0.24 & 6.50 & 0.15 & 1.78 & 0.016 &   -- & (--) &   15 &  0 \\ 
OC-0010 & 272.46 &   0.01 & -23.65 &   0.02 &   7.29 &   0.02 &  -2.05 &   0.01 &   0.02 &   0.79 &   0.02 &   1.89 &   0.19 &  -1.47 &   0.25 & 6.00 & 0.14 & 1.44 & 0.022 &   -- & (--) &   16 &  0 \\ 
OC-0011 & 272.99 &   0.05 & -22.97 &   0.05 &   8.12 &   0.05 &  -2.15 &   0.04 &   0.06 &   0.61 &   0.05 &   0.45 &   0.17 &   0.81 &   0.19 & 9.10 & 0.21 & 3.78 & 0.020 &  11.41 &   4.88 &  289 &  6 \\ 
OC-0012 & 269.75 &   0.03 & -20.59 &   0.02 &   8.73 &   0.02 &   1.64 &   0.02 &   0.03 &   0.32 &   0.01 &  -0.44 &   0.13 &  -1.83 &   0.10 & 8.00 & 0.18 & 4.04 & 0.028 &   -- & (--) &   27 &  0 \\ 
OC-0013 & 271.22 &   0.04 & -21.36 &   0.06 &   8.74 &   0.06 &   0.06 &   0.04 &   0.07 &   0.54 &   0.02 &   0.28 &   0.28 &  -1.03 &   0.36 & 7.60 & 0.17 & 4.04 & 0.028 &   -- & (--) &   18 &  0 \\ 
OC-0014 & 271.62 &   0.03 & -21.40 &   0.03 &   8.88 &   0.03 &  -0.28 &   0.03 &   0.04 &   0.49 &   0.05 &  -0.37 &   0.14 &  -1.50 &   0.17 & 9.30 & 0.21 & 4.04 & 0.028 &  36.95 &   (--) &   79 &  1 \\ 
OC-0015 & 271.27 &   0.05 & -21.21 &   0.02 &   8.89 &   0.03 &   0.10 &   0.04 &   0.05 &   0.33 &   0.01 &   0.34 &   0.08 &  -1.29 &   0.12 & 6.80 & 0.16 & 3.68 & 0.016 &   -- & (--) &   23 &  0 \\ 
OC-0016 & 270.23 &   0.04 & -20.21 &   0.03 &   9.29 &   0.03 &   1.44 &   0.03 &   0.04 &   0.38 &   0.01 &   0.19 &   0.07 &  -0.11 &   0.07 & 6.10 & 0.14 & 2.14 & 0.020 &   -- & (--) &   15 &  0 \\ 
OC-0017 & 271.95 &   0.04 & -19.50 &   0.06 &  10.69 &   0.06 &   0.38 &   0.04 &   0.07 &   0.35 &   0.01 &  -0.53 &   0.15 &  -1.96 &   0.19 & 9.80 & 0.23 & 2.20 & 0.016 &   -- & (--) &   23 &  0 \\ 
OC-0018 & 272.03 &   0.02 & -19.41 &   0.02 &  10.81 &   0.02 &   0.36 &   0.03 &   0.03 &   0.54 &   0.02 &   1.00 &   0.19 &  -1.70 &   0.24 & 6.70 & 0.15 & 4.04 & 0.026 &  -2.40 &   (--) &   16 &  1 \\ 
OC-0019 & 274.12 &   0.03 & -20.14 &   0.02 &  11.11 &   0.02 &  -1.72 &   0.02 &   0.03 &   0.52 &   0.02 &   1.17 &   0.20 &  -0.31 &   0.17 & 7.50 & 0.17 & 2.28 & 0.028 &   -- & (--) &   15 &  0 \\ 
OC-0020 & 273.46 &   0.02 & -18.99 &   0.04 &  11.83 &   0.04 &  -0.63 &   0.03 &   0.05 &   0.53 &   0.01 &   0.26 &   0.13 &  -1.64 &   0.16 & 7.30 & 0.17 & 1.80 & 0.027 &   -- & (--) &   15 &  0 \\ 
OC-0021 & 274.36 &   0.03 & -19.09 &   0.04 &  12.14 &   0.04 &  -1.42 &   0.03 &   0.05 &   0.55 &   0.03 &   0.64 &   0.11 &  -2.43 &   0.10 & 7.20 & 0.17 & 2.68 & 0.028 &   -- & (--) &   33 &  0 \\ 
OC-0022 & 274.62 &   0.03 & -18.39 &   0.02 &  12.88 &   0.02 &  -1.31 &   0.02 &   0.03 &   0.52 &   0.02 &  -0.02 &   0.25 &  -1.94 &   0.21 & 7.60 & 0.17 & 1.66 & 0.028 &   -- & (--) &   17 &  0 \\ 
OC-0023 & 273.76 &   0.02 & -17.02 &   0.02 &  13.70 &   0.02 &   0.07 &   0.02 &   0.03 &   0.35 &   0.03 &   1.21 &   0.09 &  -1.04 &   0.17 & 8.50 & 0.20 & 4.04 & 0.028 &   -- & (--) &   39 &  0 \\ 
OC-0024 & 274.18 &   0.03 & -17.01 &   0.04 &  13.89 &   0.03 &  -0.28 &   0.03 &   0.05 &   0.58 &   0.02 &   0.79 &   0.14 &  -1.90 &   0.18 & 6.80 & 0.16 & 3.26 & 0.019 &   -- & (--) &   19 &  0 \\ 
OC-0025 & 273.14 &   0.03 & -16.46 &   0.03 &  13.90 &   0.03 &   0.86 &   0.03 &   0.04 &   0.52 &   0.01 &   0.52 &   0.26 &   0.45 &   0.14 & 8.10 & 0.19 & 1.94 & 0.028 &   -- & (--) &   16 &  0 \\ 
OC-0026 & 274.94 &   0.02 & -16.54 &   0.03 &  14.66 &   0.03 &  -0.70 &   0.02 &   0.04 &   0.57 &   0.01 &   0.10 &   0.15 &  -1.77 &   0.15 & 6.50 & 0.15 & 3.98 & 0.016 &   -- & (--) &   16 &  0 \\ 
OC-0027 & 274.95 &   0.03 & -15.71 &   0.04 &  15.39 &   0.03 &  -0.32 &   0.03 &   0.05 &   0.44 &   0.01 &  -0.10 &   0.10 &  -2.06 &   0.11 & 7.40 & 0.17 & 3.58 & 0.018 &   -- & (--) &   32 &  0 \\ 
OC-0028 & 274.64 &   0.04 & -15.19 &   0.04 &  15.71 &   0.04 &   0.19 &   0.04 &   0.05 &   0.57 &   0.03 &  -0.53 &   0.14 &  -0.89 &   0.13 & 7.10 & 0.16 & 3.50 & 0.026 &  -6.80 &   (--) &   41 &  1 \\ 
OC-0029 & 275.46 &   0.04 & -13.89 &   0.04 &  17.23 &   0.04 &   0.11 &   0.03 &   0.05 &   0.48 &   0.05 &  -0.64 &   0.20 &  -2.25 &   0.19 & 6.00 & 0.14 & 3.12 & 0.028 &   -- & (--) &  121 &  0 \\ 
OC-0030 & 273.80 &   0.04 & -12.56 &   0.03 &  17.64 &   0.03 &   2.16 &   0.04 &   0.05 &   0.61 &   0.03 &  -0.47 &   0.24 &  -0.79 &   0.33 & 7.10 & 0.16 & 2.84 & 0.028 &   -- & (--) &   19 &  0 \\ 
OC-0031 & 276.29 &   0.02 & -13.28 &   0.02 &  18.15 &   0.02 &  -0.32 &   0.02 &   0.03 &   0.49 &   0.02 &  -0.45 &   0.12 &  -2.32 &   0.10 & 6.00 & 0.14 & 2.74 & 0.028 &   -- & (--) &   50 &  0 \\ 
OC-0032 & 277.26 &   0.04 & -12.71 &   0.04 &  19.08 &   0.04 &  -0.89 &   0.03 &   0.05 &   0.52 &   0.01 &   0.64 &   0.15 &  -1.64 &   0.17 & 7.00 & 0.16 & 2.48 & 0.017 &   -- & (--) &   25 &  0 \\ 
OC-0033 & 276.77 &   0.03 & -12.03 &   0.02 &  19.47 &   0.02 &  -0.16 &   0.03 &   0.04 &   0.37 &   0.02 &  -0.32 &   0.09 &  -2.98 &   0.10 & 7.50 & 0.17 & 3.26 & 0.027 &  37.64 &   (--) &   47 &  1 \\ 
OC-0034 & 279.00 &   0.03 & -12.80 &   0.02 &  19.79 &   0.02 &  -2.44 &   0.03 &   0.03 &   0.53 &   0.02 &   0.33 &   0.14 &  -0.51 &   0.16 & 7.60 & 0.17 & 1.98 & 0.028 &   3.65 &   (--) &   19 &  1 \\ 
OC-0035 & 276.84 &   0.03 & -10.96 &   0.04 &  20.45 &   0.03 &   0.28 &   0.04 &   0.05 &   0.31 &   0.04 &  -0.43 &   0.11 &  -2.06 &   0.14 & 9.40 & 0.22 & 3.20 & 0.016 &   7.16 &   1.73 &   39 &  2 \\ 
OC-0036 & 277.18 &   0.05 & -10.40 &   0.05 &  21.10 &   0.05 &   0.25 &   0.05 &   0.07 &   0.31 &   0.01 &  -0.56 &   0.15 &  -2.41 &   0.18 & 9.80 & 0.23 & 2.28 & 0.017 &   -- & (--) &   33 &  0 \\ 
OC-0037 & 278.09 &   0.03 & -10.57 &   0.03 &  21.36 &   0.03 &  -0.63 &   0.03 &   0.04 &   0.52 &   0.02 &   0.65 &   0.17 &  -0.20 &   0.12 & 7.70 & 0.18 & 4.04 & 0.028 & -10.36 &   (--) &   22 &  1 \\ 
OC-0038 & 279.22 &   0.06 & -10.91 &   0.03 &  21.57 &   0.04 &  -1.76 &   0.05 &   0.06 &   0.53 &   0.01 &  -0.95 &   0.22 &  -2.72 &   0.32 & 6.90 & 0.16 & 4.04 & 0.028 &   -- & (--) &   20 &  0 \\ 
OC-0039 & 279.29 &   0.02 & -10.24 &   0.02 &  22.20 &   0.02 &  -1.52 &   0.03 &   0.03 &   0.32 &   0.01 &  -1.45 &   0.10 &  -4.33 &   0.16 & 9.00 & 0.21 & 3.68 & 0.028 &   -- & (--) &   19 &  0 \\ 
OC-0040 & 277.95 &   0.03 &  -9.15 &   0.05 &  22.56 &   0.04 &   0.15 &   0.03 &   0.05 &   0.33 &   0.01 &  -0.83 &   0.23 &  -3.18 &   0.35 & 9.80 & 0.23 & 1.94 & 0.017 &   -- & (--) &   19 &  0 \\
\multicolumn{19}{c}{\hspace{3.3cm}  ... } \\ \\ \hline \\ \\
\end{tabular}
\end{table*}

\begin{figure*}
\centering
\includegraphics[width=0.325\linewidth]{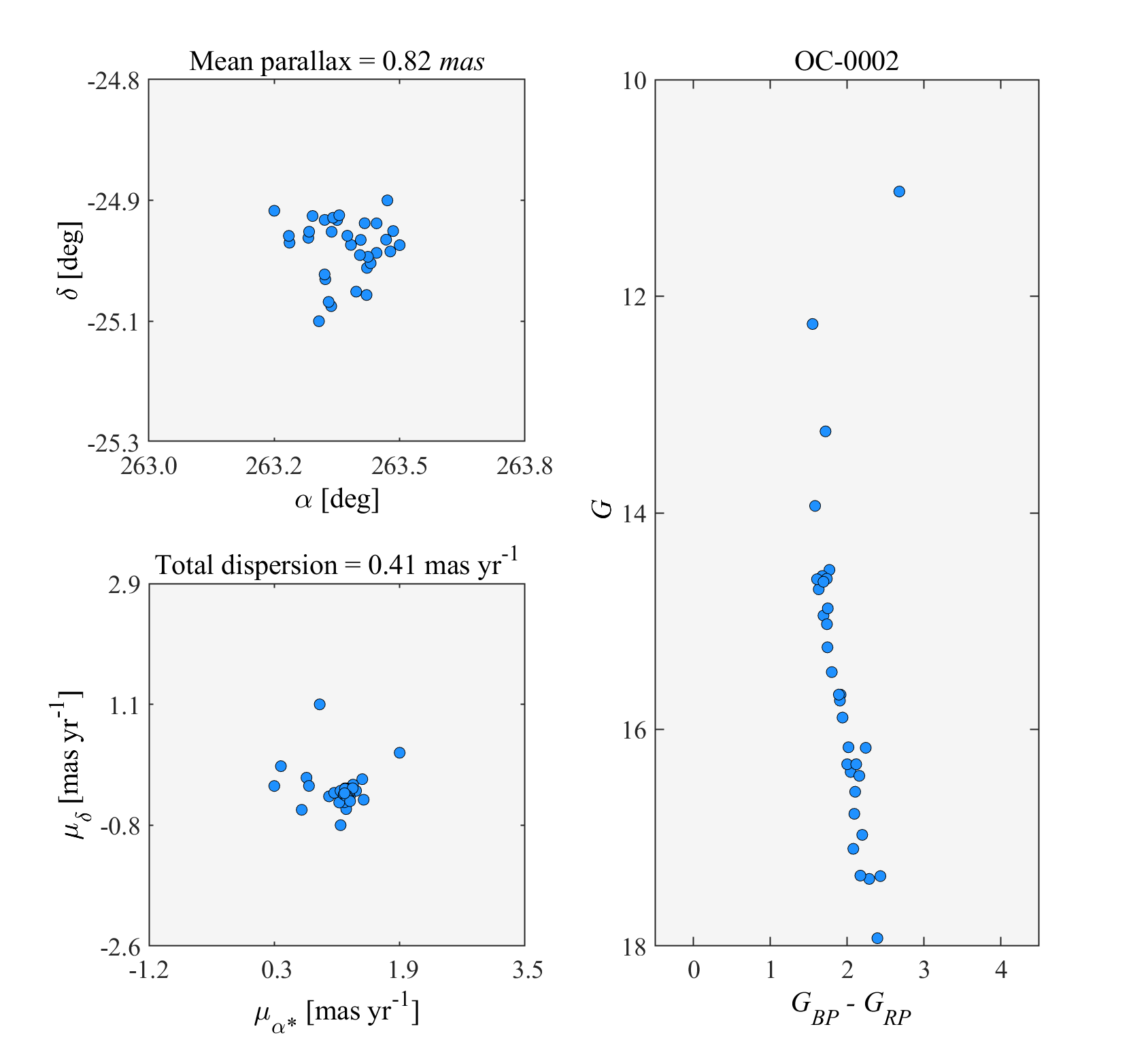} \hspace{0.0cm}
\includegraphics[width=0.325\linewidth]{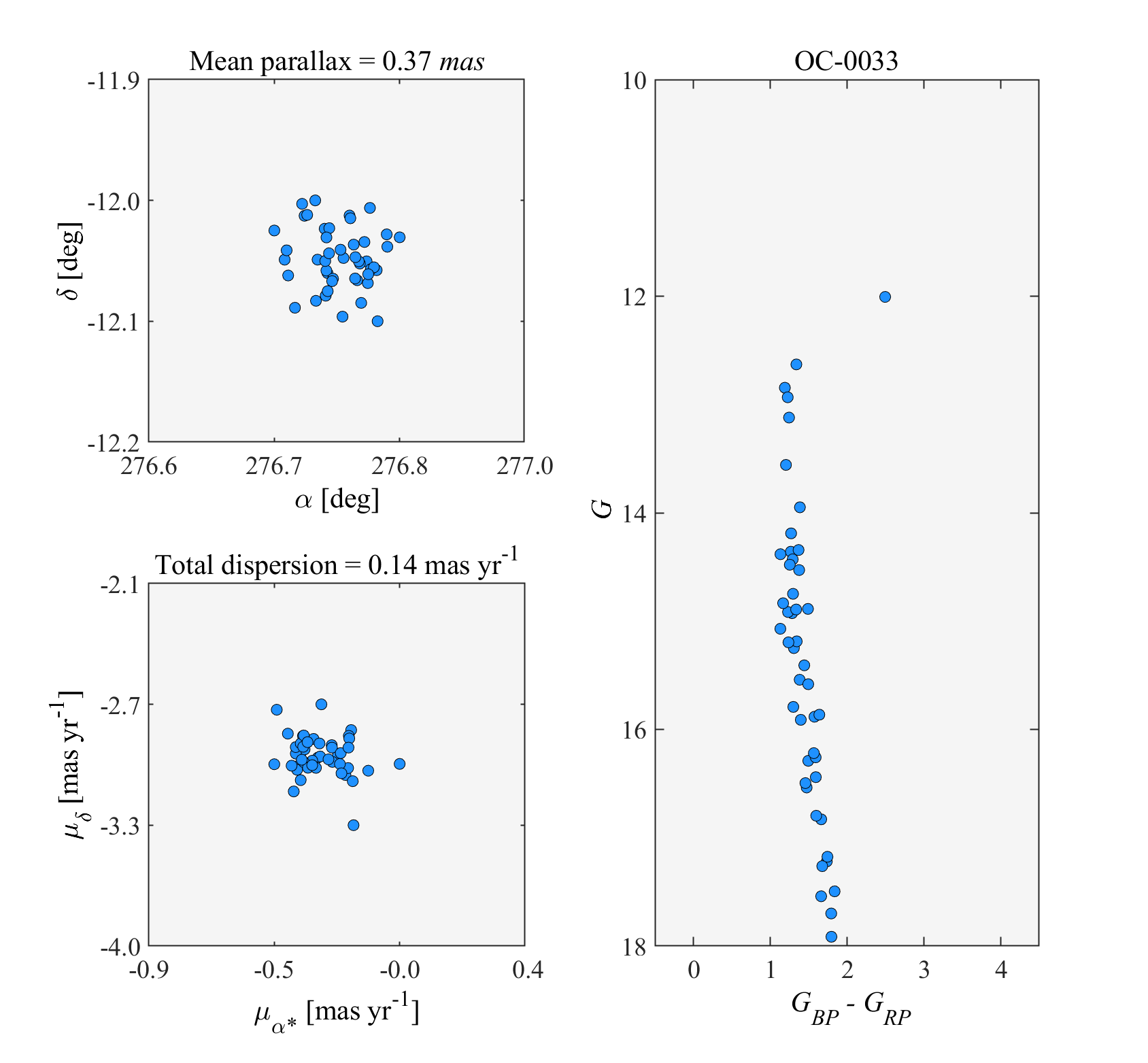}  \hspace{0.0cm}
\includegraphics[width=0.325\linewidth]{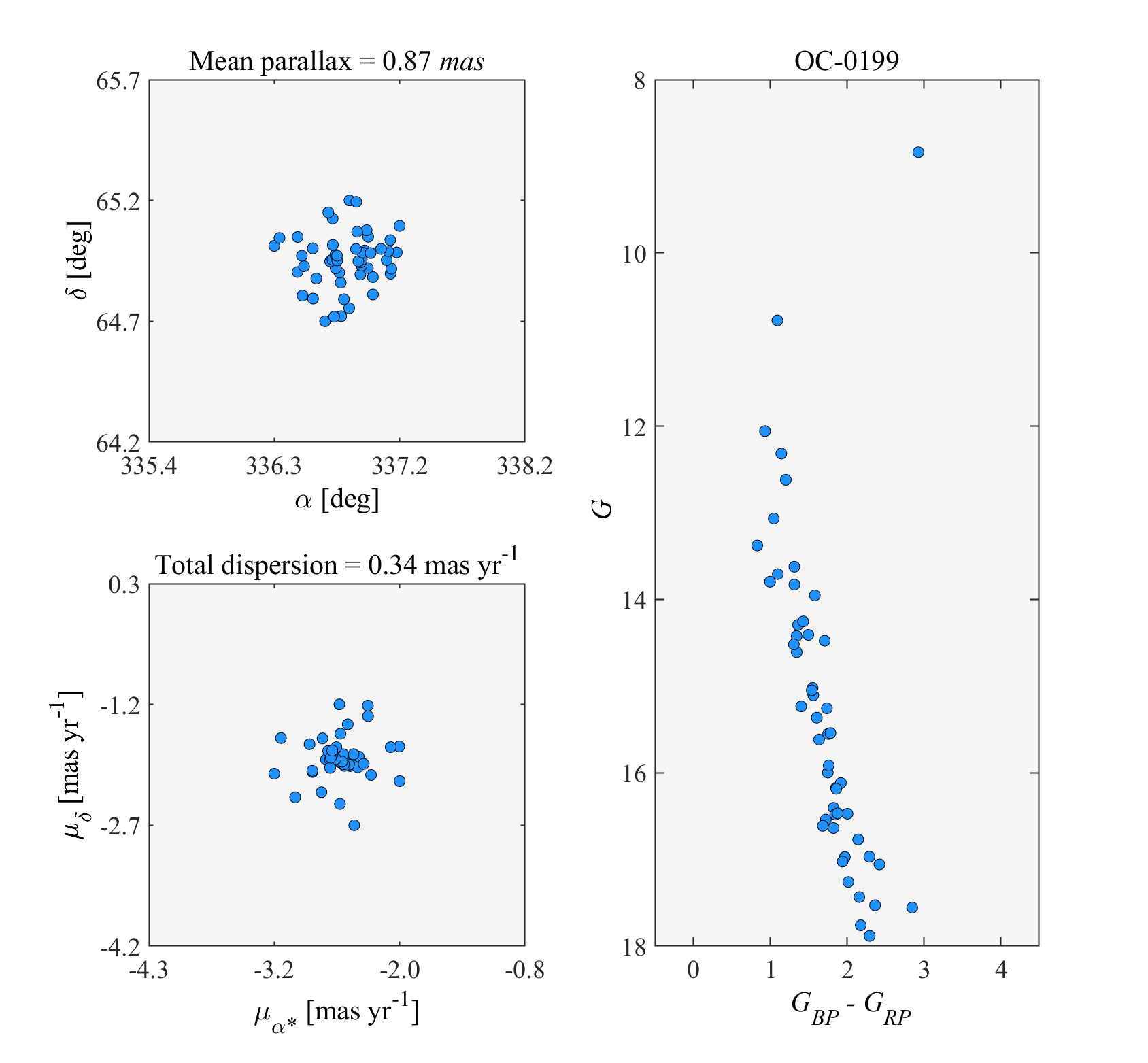} \hspace{0.0cm}
\includegraphics[width=0.325\linewidth]{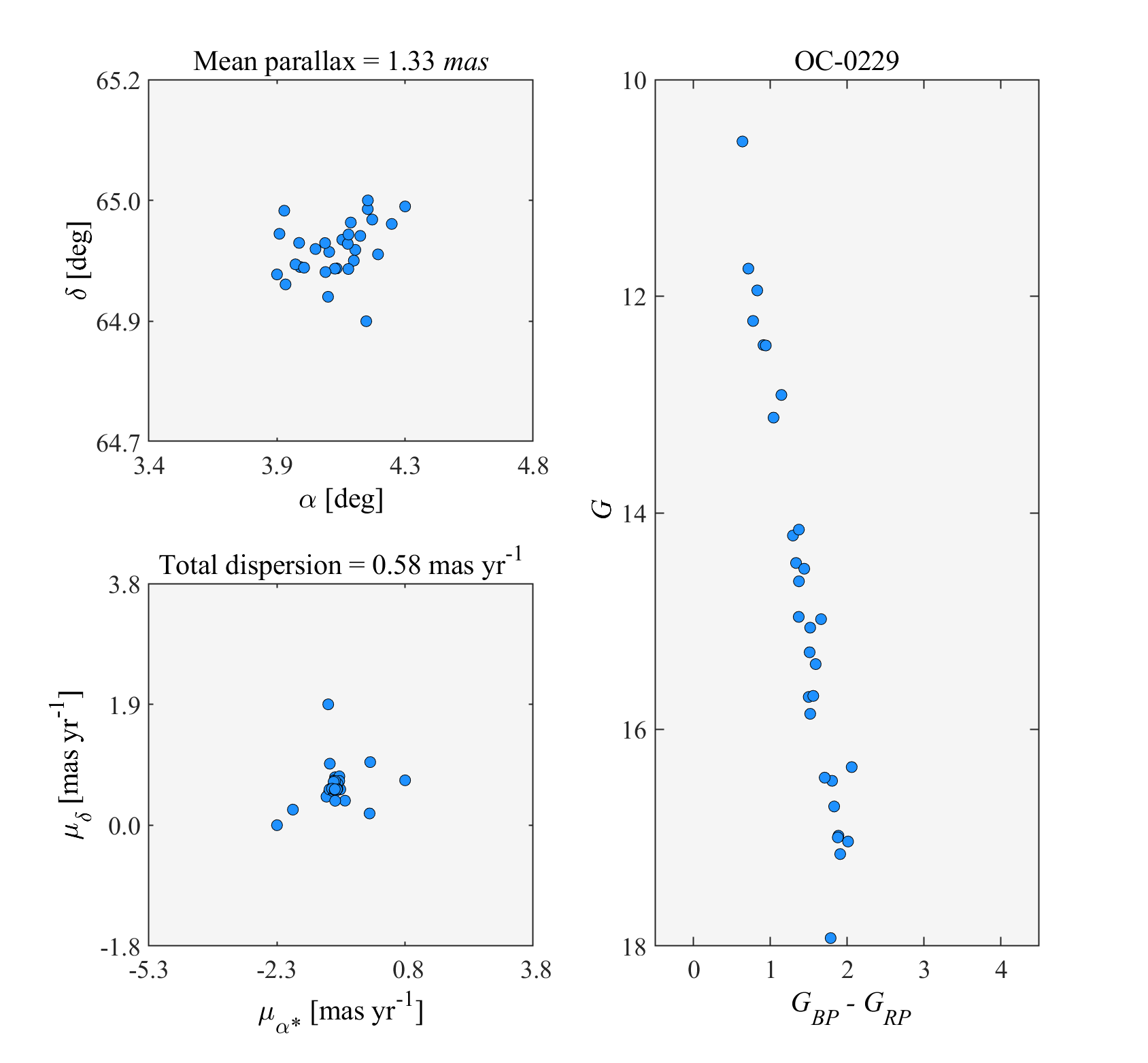} \hspace{0.0cm}
\includegraphics[width=0.325\linewidth]{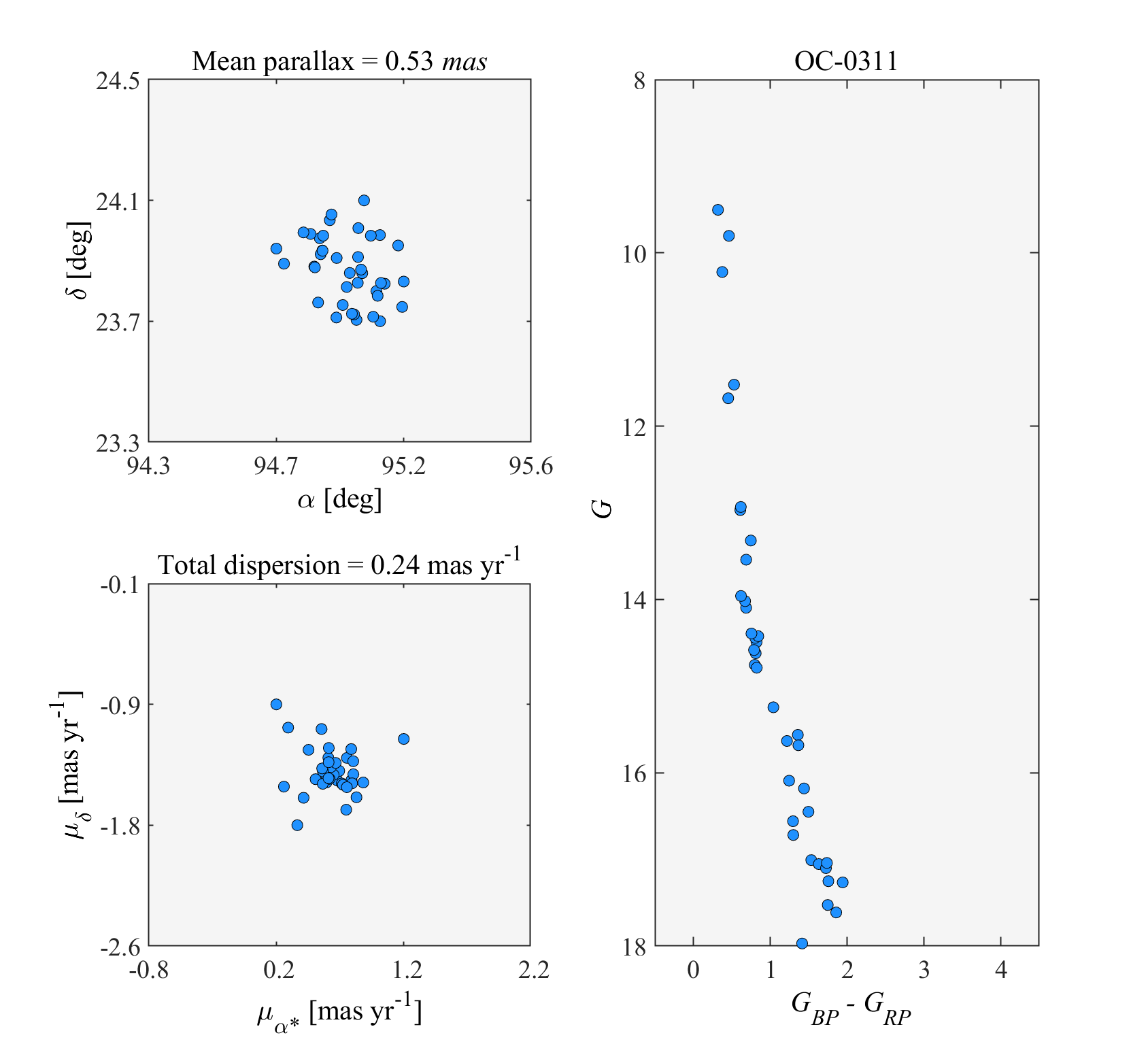}  \hspace{0.0cm}
\includegraphics[width=0.325\linewidth]{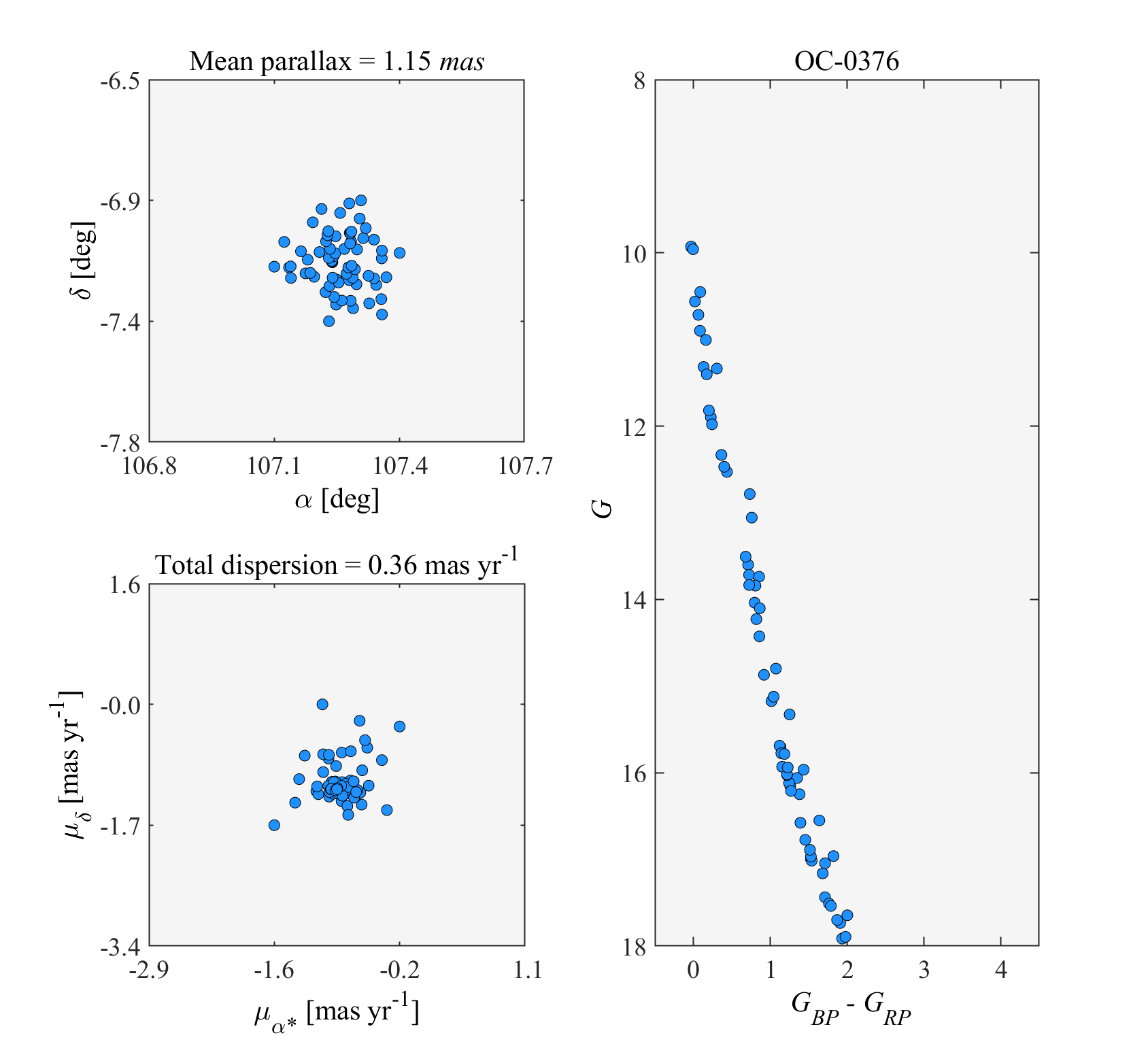} \hspace{0.0cm}
\includegraphics[width=0.325\linewidth]{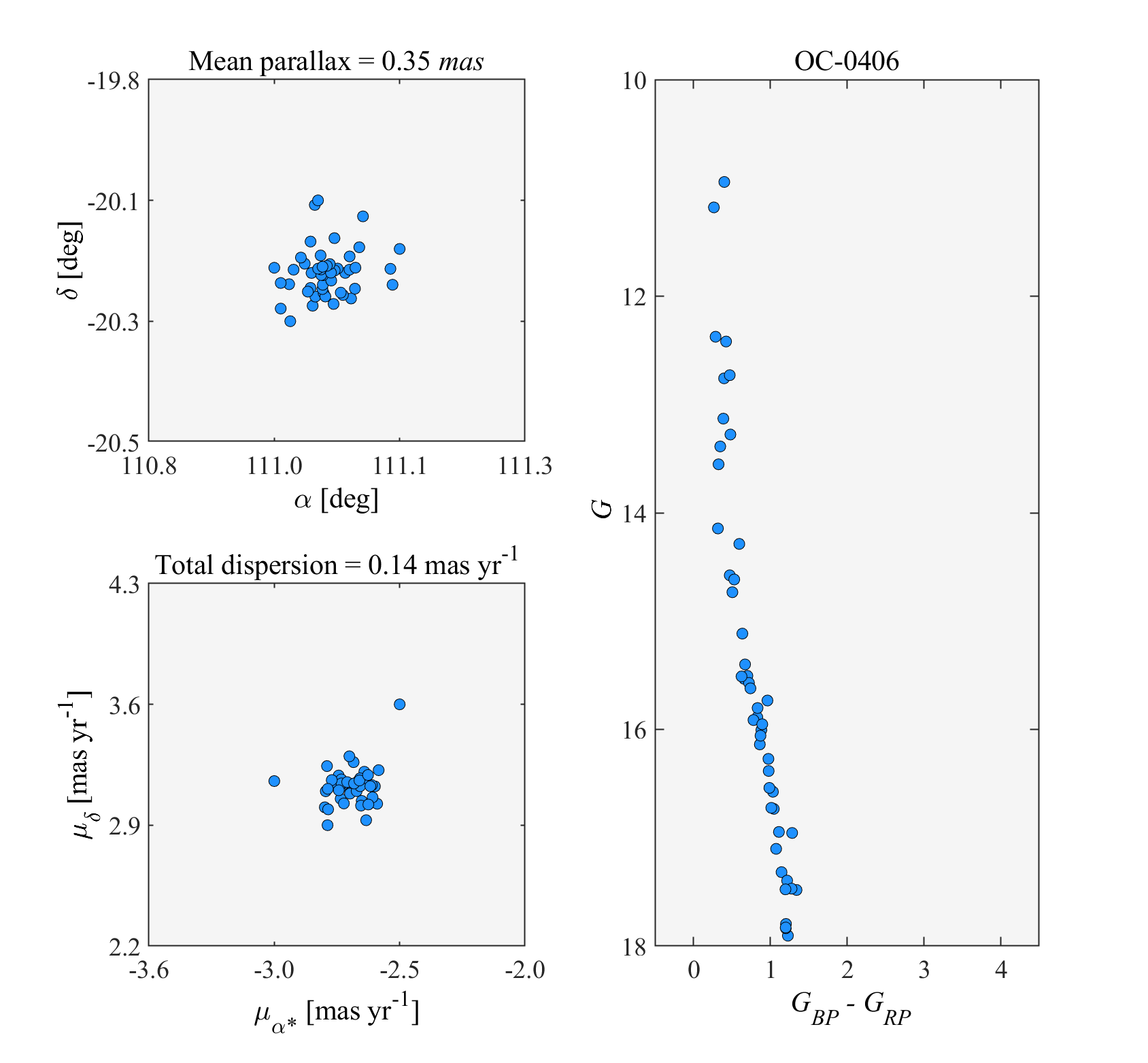}  \hspace{0.0cm}
\includegraphics[width=0.325\linewidth]{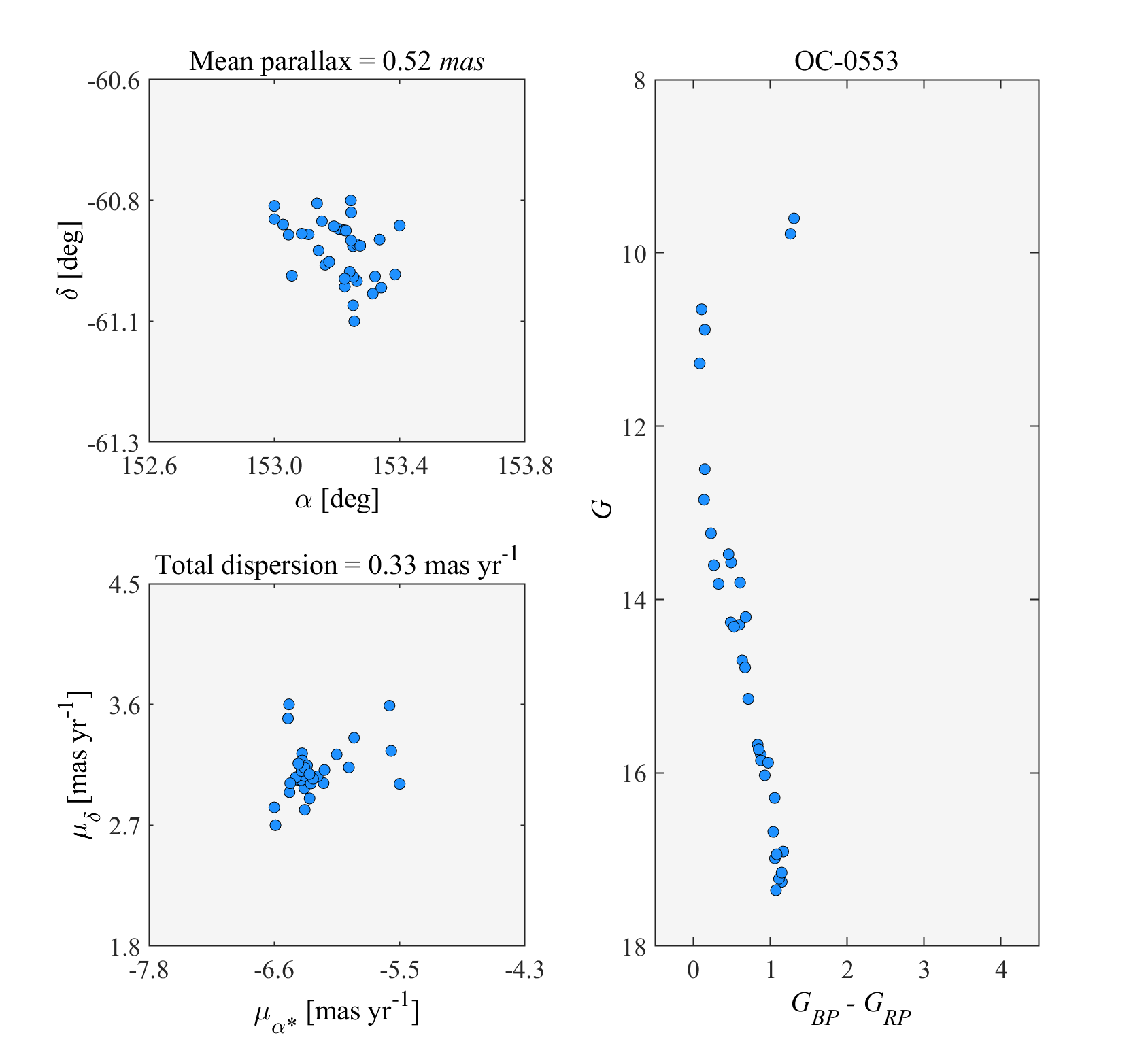} \hspace{0.0cm}
\includegraphics[width=0.325\linewidth]{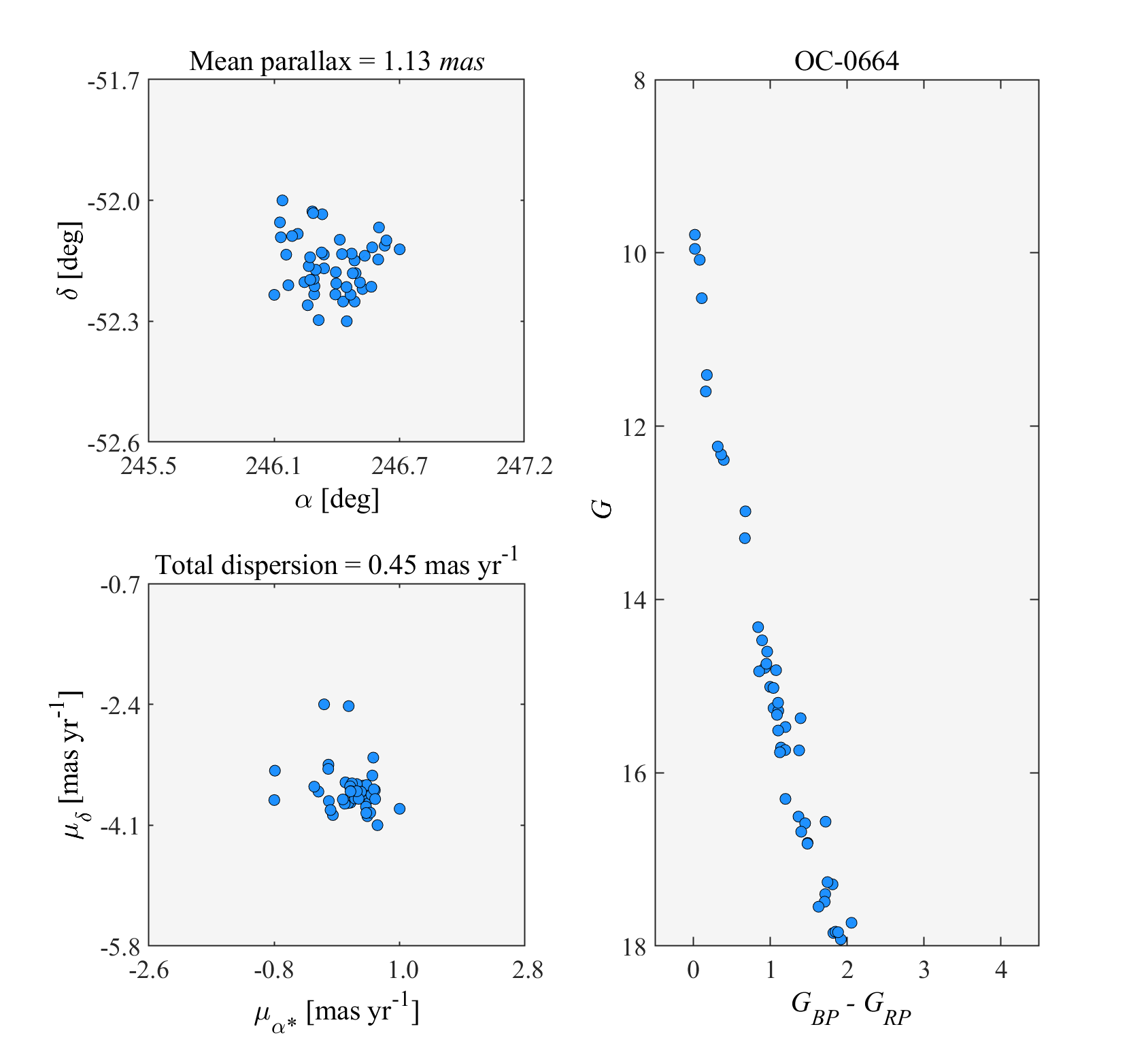} \hspace{0.0cm}
\caption{Examples of class I and class II OCs. For each OC, the columns represent the distributions of the member stars for position in RA and Dec, proper motions in $\mu_{\alpha^{*}}$ and $\mu_{\delta}$, and the CMD in $G$ vs. $G_{\rm BP}$ - $G_{\rm RP}$, including its mean parallax and total proper-motion
dispersions. Here, the listed OCs are OC-0002, OC-0033, OC-0199, OC-0229, OC-0311, OC-0376 and OC-0406, OC-0553, and OC-0664.}
\label{fig:new_OCs}
\end{figure*}


\begin{figure}
\centering
\includegraphics[width=1.02\linewidth]{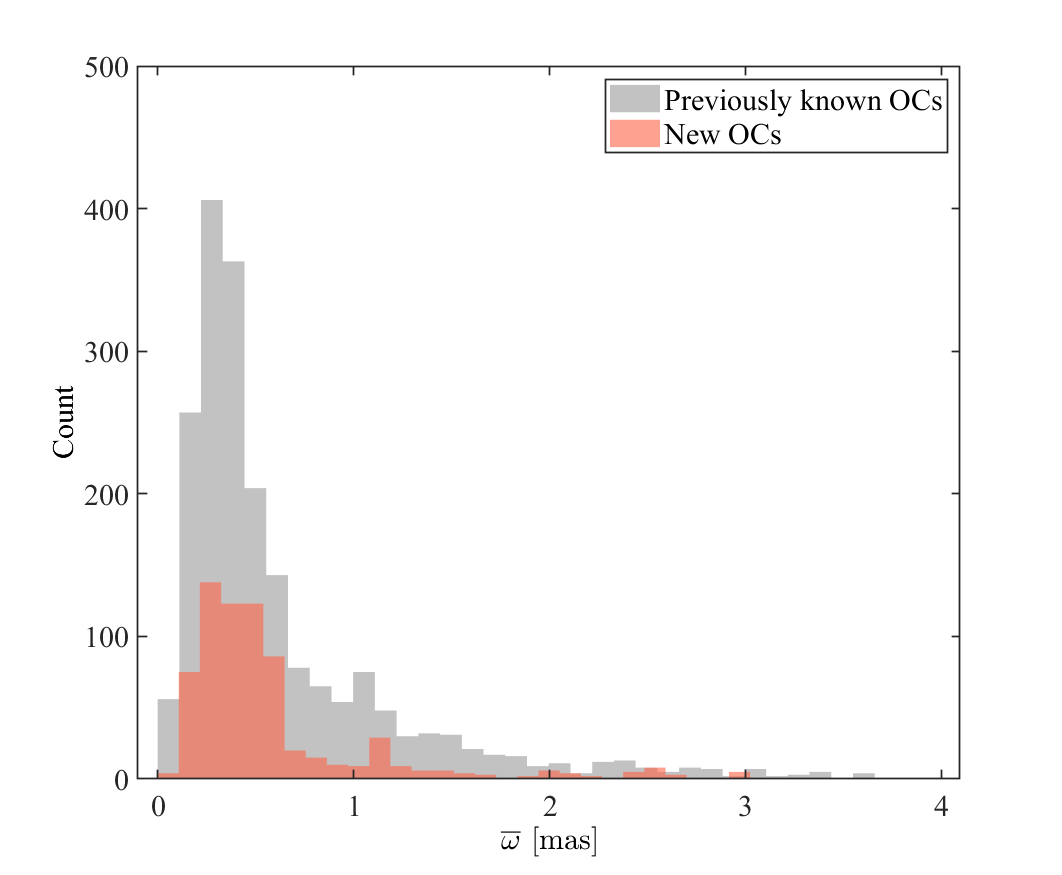} \hspace{0.0cm}
\caption{Parallactic distributions of the newly found OCs (orange) in this work and previously known OCs (grey) from \cite{cantat2020b}.}
\label{fig:parallax}
\end{figure}

\subsection{Clustering algorithm}
\label{algorithm}
After preparing the samples (as described in Sect.~\ref{sample}) we 
began to search for spatial overdensity structures using the unsupervised
clustering algorithm, DBSCAN. This is a density-based algorithm, 
so it offers advantages that allow it to find clusters of arbitrary shapes. 
As the member stars in an OC have common positions and proper motions, 
we applied DBSCAN in the five-dimensional (5D) space ($l$, $b$, $\varpi$, 
$\mu_{\alpha^{*}}$ and $\mu_{\delta}$). The DBSCAN algorithm defines a 
cluster based on two parameters: $\epsilon$ and $minPts$, which are used
to describe the closeness of the sample distribution in the neighbourhood. 
The parameter $\epsilon$ is automatically computed in each sample using 
a Gaussian kernel density estimation method and the $k$-nearest 
neighbours algorithm (more details see Paper I). 
In this work, we were still focused on smaller regions as Paper I to 
detect unnoticed OCs, namely, the sizes of the squares in different distance 
bins were set in the range of [$1^{\circ}$, $10^{\circ}$]. 
Meanwhile, the parameter $minPts$ is the minimum number of objects 
within $\epsilon$ that can be considered as a cluster, and since it was been 
shown in Paper I that many hidden OCs could be hunted with $minPts$ = 
[6, 10] in small regions, we adopted the same values in this work.

In addition, for stellar clusters located at the borders of the boxes in 
Sect.~\ref{sample}, we took the following measures to dispose of them. If 
there were two or more clusters with compatible mean parameters within 
3$\sigma_{i}$ (where $\sigma$ is the standard deviation) in the 
5D parameter space, $i$ = {$l$, $b$, $\varpi$, 
$\mu_{\alpha^{*}}$ and $\mu_{\delta}$}, they were considered as 
substructures of the same star cluster. After further confirmation by visual 
inspection, we combined these substructures or re-detected the entire 
cluster. In the following steps, we introduce how we determined if the 
overdensity structures found with DBSCAN were potentially real OCs or 
just statistical clusters.


\subsection{Proper-motion dispersion}
\label{dispersion}

As mentioned by \cite{cantat2020a}, while the apparent proper-motion 
dispersion of an OC does not constitute an accurate diagnosis of its 
dynamical state, it can serve as a sufficient empirical basis for distinguishing
reasonable OCs from implausible ones. 
Hence, we filtered the searched star clusters in Sect.~\ref{algorithm} by 
their internal proper-motion dispersions.
Recently, we found that this method was also adopted in \cite{hunt2021}. 
In this study, we adopted more rigorous dispersion parameters.

An OC, as a gravitationally bound system, has a small velocity dispersion. 
The typical dispersion of a globular cluster (GC) is from 5 to 10~km~s$^{-1}$
\citep[e.g.,][]{lapenna2015,baumgardt2018}. In turn,  OCs have smaller 
dispersions. For example, the study by \cite{mermilliod2009} showed that
the typical dispersion of an OC is $\sim$ 1~km~s$^{-1}$. However, recent 
studies, using line-of-sight velocities obtained from high-resolution 
spectroscopy, indicated that an OC has an internal dispersion below 
$\sim$ 2~km~s$^{-1}$ \citep[e.g.,][]{donati2014,cantat2014,vereshchagin2016,
overbeek2017,hatzidimitriou2019}. Interestingly, a recent study that focused 
on the relationship between the total proper-motion dispersions and 
parallaxes of OCs identified in \emph{Gaia} DR2 \citep{cantat2020a} showed 
that the dispersions of most OCs more distant than $\sim$ 1~kpc (i.e., 
parallax smaller than 1~mas) are larger than 2~km~s$^{-1}$ but below 
0.5~mas~yr$^{-1}$, which may result from measurement uncertainties or 
uncertainties associated with the difficulty of distinguishing between OC 
member stars and contaminating field stars. 
Thus, we selected the OC candidates using the proper-motion criterion:
\begin{equation} 
\sqrt{\sigma_{\mu_{\alpha^{*}}}^{2}+\sigma_{\mu_{\delta}}^{2}} \leq 
\begin{cases}
0.5 \ \rm mas \ yr^{-1} & \text{if \ $\varpi <\ 1 \ \rm mas$}\\
2\sqrt{2} \frac{\varpi}{4.7404} \ \rm mas \ yr^{-1} & \text{if \ $\varpi \geq \ 1 \ \rm mas$}
\end{cases}
.\end{equation}


\subsection{Confirmation}
\label{confirm}
For the OC candidates obtained in Sect.~\ref{dispersion}, here we go on to 
introduce how we proposed them as potentially real OCs. 
On the one hand, we made references to the characteristics of known 
OCs described in some previous studies~\citep[e.g.,][]{cantat2020a,cantat2020b,
tarricq2021,dias2021}; for instance, an OC candidate can be considered as 
a potentially real OC if it could be shown to clearly pass its members, distance, 
age, or velocity relative to the field stars, as well as the density of the 
background stellar distribution, for instance.
Furthermore, the member stars of an OC formed in the same molecular cloud, 
so its color--magnitude diagram (CMD) should follow an empirical isochrone. 
Hence, we visually inspected the CMDs of the OC candidates found in this 
work in order to propose the more reliable ones.

In addition, if an OC candidate appeared to be potentially real, we estimated 
its \textit{RV} and error as well as its age for further research. As for the 
weighted standard deviation of its \textit{RV}, we adopted the same equation 
described by \cite{soubiran2018}:
\begin{equation} 
\sigma_{\rm \textit{RV}_{OC}}^{2} = \frac{\underset{i}{\Sigma}w_{i}}{(\underset{i}{\Sigma}w_{i})^{2}
- \underset{i}{\Sigma}w_{i}^{2}} \underset{i}{\Sigma}w_{i}(\rm \textit{RV}_{\it i} - \textit{RV}_{OC})^{2},
\end{equation}
where $RV_{\rm OC}$ is the mean value of the member stars, $w_{i}$ =
1/$\varepsilon_{i}^{2}$ refers to the individual radial velocity measurements
and $\varepsilon_{i}$ is the \textit{RV} error for star $i$ provided in 
\textit{Gaia} EDR3. 

Many efforts have been devoted to determining the ages of OCs in 
\textit{Gaia }\citep[e.g.,][]{Bossini2019,cantat2020b}.
We also determined the ages of the proposed new OCs using the same 
method adopted in some previous studies \citep[e.g.,][]{liu2019,hao2020,
he2021}, which is aimed at constructing the best isochrone for the member 
stars in an OC candidate. 
The isochrones are derived from the PARSEC library \citep{bressan2012} 
updated for the \textit{Gaia} DR2 passbands with the photometric 
calibration of \cite{evans2018}.
These isochrones we used contain logarithmic ages (i.e., log(age/yr)) from 
5.92 to 10.13 and metal fractions ($z$) from 0.015 to 0.029, as well as a series 
of isochrones are generated with steps of $\Delta$ log(age/yr) = 0.02 and 
$\Delta$ $z$ = 0.001, respectively.

The fitting function is the key to constructing a reliable isochrone for an OC
and isochrone fitting pipelines have been produced in several studies 
\citep[e.g.,][]{Perren2015,Bonatto2019}.
As in some previous works, e.g., \cite{liu2019} and \cite{he2021},
we used the following fitting function to fit the CMDs for OCs:
\begin{equation} 
\centering
\overline{d}\ ^{2} = \sum_{k=1}^{n} | \ \textbf{\emph{x}}_{k} - \textbf{\emph{x}}_{k, nn} \ |^{2} / n,
\end{equation}
where $n$ is the number of member stars in an OC candidate, and 
$\textbf{\emph{x}}_{k} = [{\it G}_{k} + \Delta_{\rm G} + {\it A}_{\rm G}, 
\ ({\it G}_{\rm BP} - {\it G}_{\rm RP})_{\it k} + {\it E}({\it G}_{\rm BP} 
- {\it G}_{\rm RP})]$, $\textbf{\emph{x}}_{k, nn}$ is the position of the 
$k$th member star and the nearest neighboring point to this star in the
isochrone, respectively.
For the $\textbf{\emph{x}}_{k}$, $\Delta_{\rm G}$ is the distance modulus,
{\it A}$_{\rm G}$ denotes the line-of-sight extinction, and 
${\it E}({\it G}_{\rm BP} - {\it G}_{\rm RP})$ indicates the reddening in $G$ 
magnitude of an OC candidate. 

The extinction {\it A}$_{\rm G}$ and reddening ${\it E}({\it G}_{\rm BP} - 
{\it G}_{\rm RP})$ of an OC candidate were corrected as follows.
When fitting isochrones, we set the parameter {\it A}$_{\rm G}$ from 0.0 
to 4.0 with a step of 0.02 and ${\it E}({\it G}_{\rm BP} - {\it G}_{\rm RP})$ 
= 0.50 {\it A}$_{\rm G}$, considering the approximate relation between 
{\it A}$_{\rm G}$ and ${\it E}({\it G}_{\rm BP} - {\it G}_{\rm RP})$ in 
\textit{Gaia} \citep{Andrae2018} as well as an extinction curve 
of $R_{v}$ = 3.1 for the Milky Way galaxy \citep{Cardelli1989,ODonnell1994}.
In addition, parameter $\overline{d}\ ^{2}$ is the mean square distance 
between OC members and their closest neighboring points in the isochrone.
For each OC candidate, we minimized $\overline{d}\ ^{2}$ by calculating 
the average squares of the distances between member stars and 
corresponding points in each theoretical isochrone.
As mentioned by \cite{liu2019}, this method is sensitive to the discrepancy 
between isochrones and the observational data.


\section{Results}
\label{sect:results}
In total, we detected thousands of stellar clusters in the Galactic disk. 
After cross-matching with the catalogs from previous works, we obtained 
2 012 known Galactic stellar clusters, including 1 930 OCs and 82 Galactic 
globular stellar clusters. Among the remaining spatial stellar clusters, we 
rejected those clusters with large proper-motion dispersions and selected 
the resulting 910  clusters as OC candidates. Finally, after inspecting the 
CMDs and other information of these OC candidates, we proposed 704 as 
potentially real OCs that had previously gone undetected.


\subsection{Cross-matching with known OCs in \textit{Gaia}}
\label{sect:gaia_oc}

For all of our detected spatial star clusters, we first cross-matched them
with known OC catalogs in \textit{Gaia} \citep[i.e.,][]{cantat2018,castro2018,cantat2019,castro2019,sim2019,liu2019,ferreira2019,
hao2020,castro2020,ferreira2020,he2021,ferreira2021,hunt2021,castro2021b}. 
If a stellar cluster found in this work and a known OC from these catalogs 
were found to have compatible mean parameters within 3$\sigma_{i}$ in 
the  5D astrometric parametric space, $i$ = {$l$, $b$, $\varpi$, 
$\mu_{\alpha^{*}}$ and $\mu_{\delta}$}, we considered this star cluster 
as cross-matched and a visual inspection was also adopted at the same 
time to confirm this assessment. 

Although our aim is to detect hidden OCs in our Galaxy, we still 
re-detected a large amount of previously known OCs in the \textit{Gaia} 
data; for instance: 889 OCs that listed in the catalog ($\sim$1 200 known OCs) 
of \cite{cantat2018} were refound; we were able to find 37 OCs in the catalog 
(41 known OCs) of \cite{cantat2019}; 647 OCs reported in \cite{castro2018,
castro2019,castro2020,castro2021b} were cross-matched; and so on.
In total, the 1 759 stellar clusters  we obtained were cross-matched with the 
above catalogs, located in the Galactic latitudes of $|b| \le$ 20$^{\circ}$, 
which is comparable with the result (1 559) in the recent work of 
\cite{castro2021b}, showing a high re-detection efficiency.
%

\begin{figure*}
\centering
\includegraphics[width=0.85\linewidth]{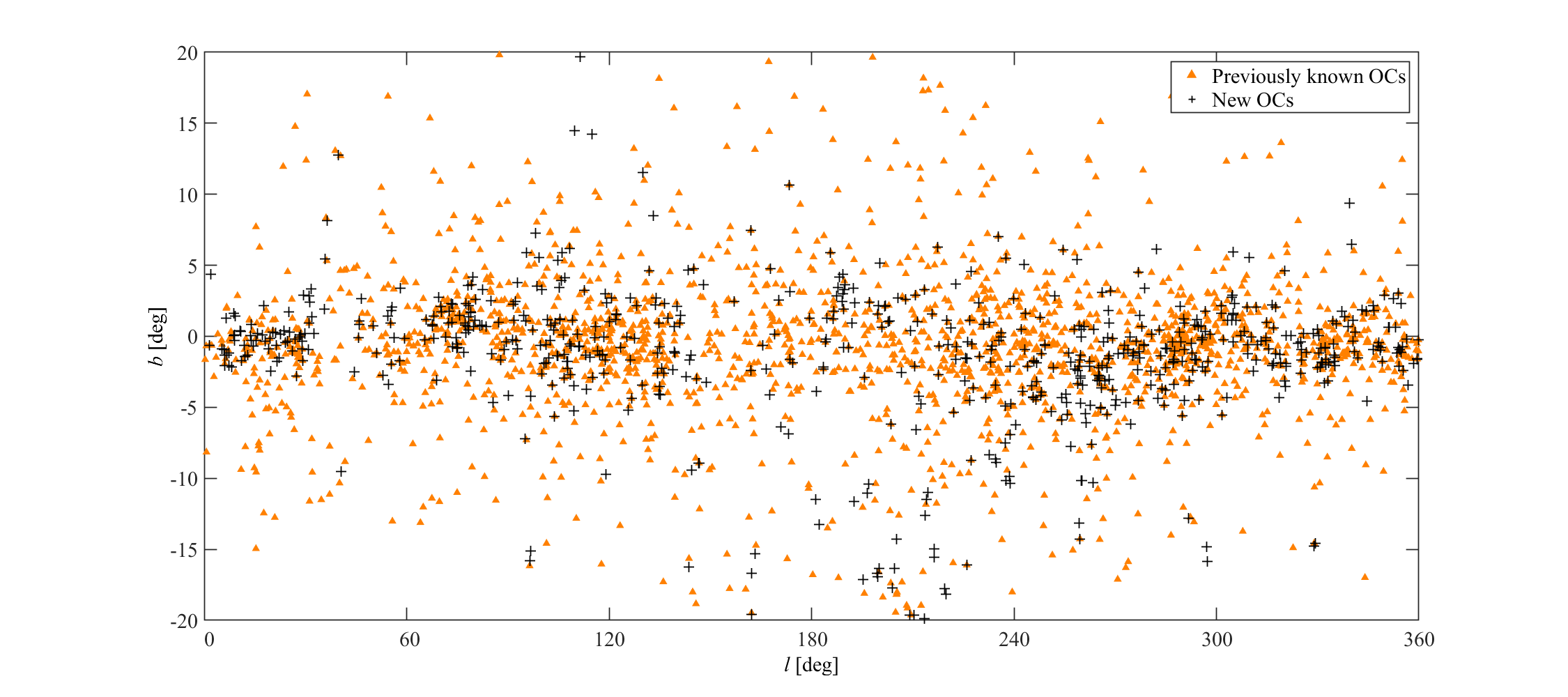} \hspace{0.0cm}
\caption{Distributions of $l$ versus $b$ of the newly found OCs and previously known OCs, where the black plus signs represent the new OCs and the orange triangles represent the known OCs compiled in \cite{cantat2020b}. }
\label{fig:lb}
\end{figure*}
\begin{figure*}
\centering
\includegraphics[width=0.86\linewidth]{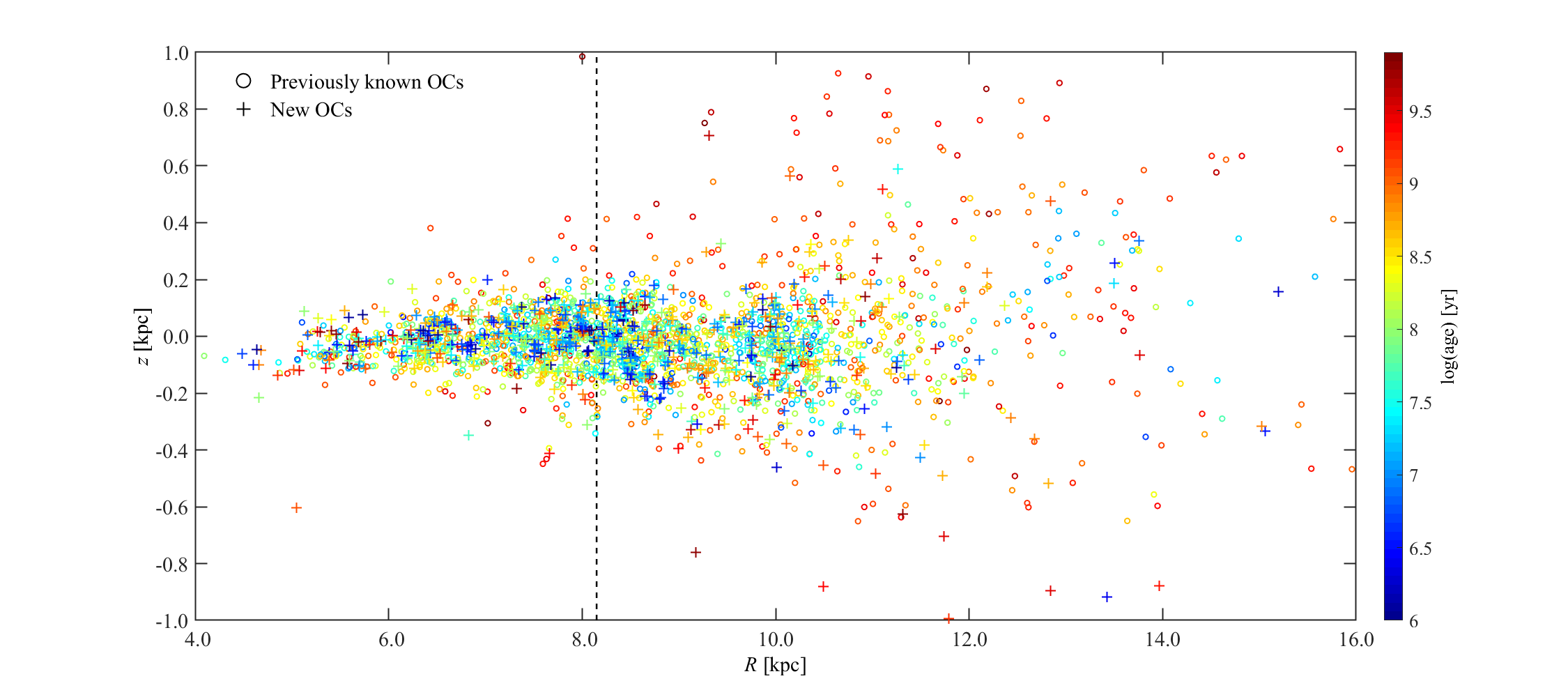} \hspace{0.0cm}
\caption{Distributions of $R$ versus $z$ of the newly found OCs and previously known OCs, and the same as Fig.~\ref{fig:lb}, the plus signs represent the new OCs and the triangles represent the known OCs (\cite{cantat2020b}), but the ages of the OCs are color-coded here. The Solar circle (black dashed) is at 8.15~kpc~\citep{reid2019}.}
\label{fig:age_r_z}
\end{figure*}

\subsection{Cross-matching with other known OCs}
\label{sect:other-oc}
Before \textit{Gaia}, \cite{dias2002} presented a list of OCs based on 
revised data compiled from previous catalogs and from isolated papers 
published before, which contained more than 2 000 objects. Using the 
near-infrared photometric data of approximately 470 million objects in 
2MASS \citep{skrutskie2006}, and proper motions from the PPMXL 
catalog \citep{roser2010}, \cite{kharchenko2013} completed a survey
of all previously known OCs, called the Milky Way Star Clusters (MWSC). 
This catalog lists 3 006 objects, including known OCs and 
OC candidates (2 469), GCs, associations, and asterisms.  

After completing the step in Sect. \ref{sect:gaia_oc},
we cross-matched the remaining clusters with the above two 
catalogs using the following criteria: if a star cluster falls within a 
circle of $0.5^{\circ}$ radii \citep{castro2019,castro2020} of one 
object listed in these two catalogs, we selected it. Then, if their 
mean proper motions were within 3$\sigma_{i}$ of the cluster found 
in this work and their distances were compatible, we considered this 
star cluster as being matched. 
The catalogs in \cite{dias2002} and \cite{kharchenko2013} do not 
provide the mean parallax and associated uncertainty for each OC 
but an estimation of the distance instead, so the criterion is that the 
known OCs are within the distance errors of our detected clusters.
The distance of our detected cluster is the inverse of the mean parallax 
of member stars~\citep{castro2020}, and the assumed errors come 
from three times the standard deviation of parallax.
Although most of the OCs in these two catalogs had already been 
taken into account by the cross-matching of our star clusters with 
the sample of \cite{cantat2018}, we still found 120 known OCs 
in the MWSC catalog and 71 known OCs in the catalog 
of \cite{dias2002}, along with 20 OCs that are repetitive.


\begin{figure*}
\centering
\includegraphics[width=0.56\linewidth]{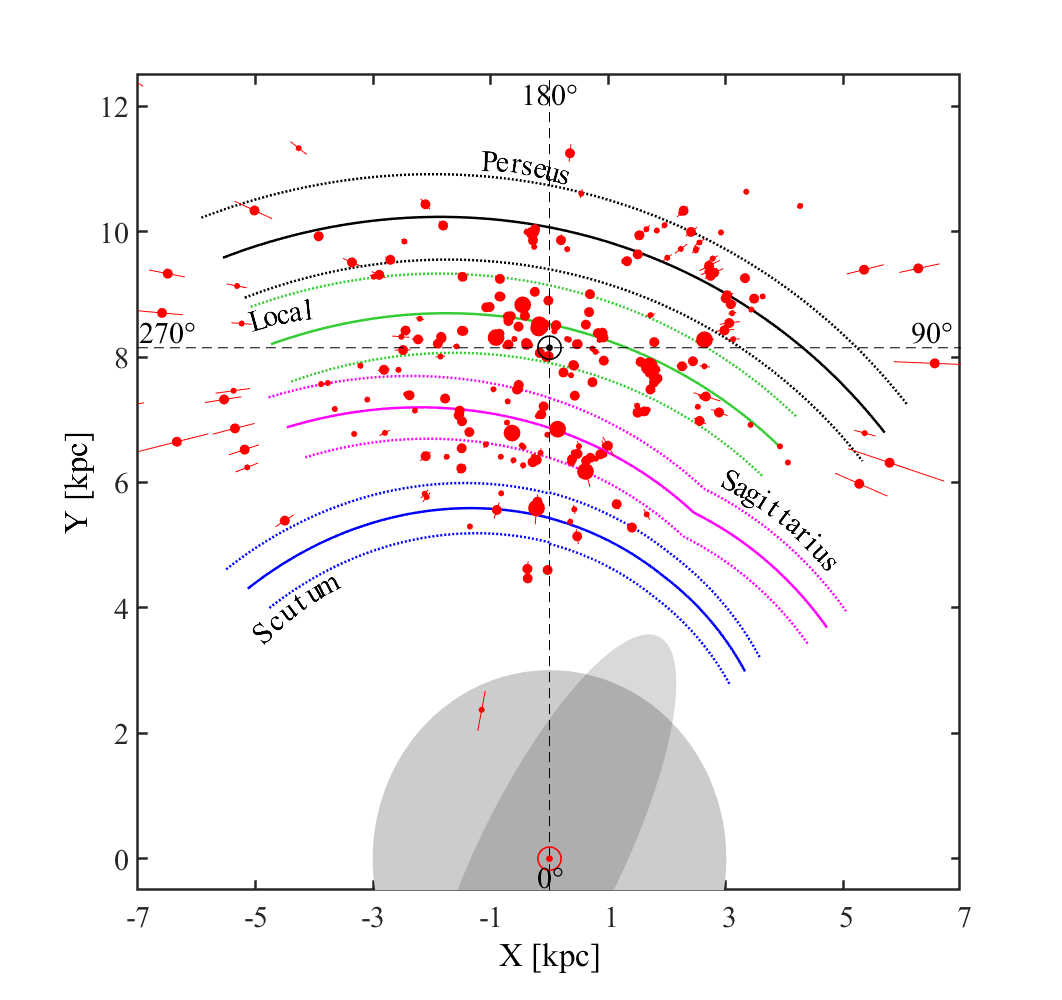} \hspace{0.0cm}
\caption{Locations of the newly found OCs younger than 20 million years projected onto the Galactic plane, including their 1$\sigma$ distance uncertainties. The sizes are proportional to the number of members of each cluster. The solid and dashed curved lines denote the arm centre and widths (i.e., which enclose 90\% of the masers), respectively, fitted by \citet[]{reid2019} from their parallax data of Very Long Baseline Interferometry masers. The Galactic Centre (red symbol) is at (0, 0) kpc and the Sun (black symbol) is at (0, 8.15)~kpc.}
\label{fig:young_ocs}
\end{figure*}


\subsection{Cross-matching with known GCs}
\label{sect:gc}
In order to cross-match with known GCs more accurately, we 
investigated the GC catalog reported by \cite{baumgardt2019}, who 
presented the mean proper motions and space velocities of 154 
Galactic GCs using a combination of \textit{Gaia} DR2 proper motions 
and ground-based line-of-sight velocities. Finally, we found that 82 star 
clusters we obtained were cross-matched with the GCs listed by 
\cite{baumgardt2019}, and the proper motions and \textit{RV}s provided 
by \textit{Gaia} EDR3 are consistent with those in the catalog.

There are 106 GCs presented by \cite{baumgardt2019} located at the 
Galactic latitude of $|b| \le $ 20$^{\circ}$, and, thus, the proportion of 
GCs detected here is $\sim$ 80\%. Each GC found here contains 
thousands of stars, and the detection of Galactic GCs is a good test 
of our proposed method. Meanwhile, for the remaining $\sim20\%$ 
of GCs that were not detected by us, their non-detection might be the
result of their far distances make their member stars being too faint.
%


\subsection{New OCs}
\label{sect:new_oc}
After cross-matching with the known Galactic OC and GC catalogs, 
there were many star clusters remaining whose cluster nature was yet 
to be determined. Among these clusters, 910 ones met the proper-motion
criterion described in Sect.~\ref{dispersion}.
These 910 star clusters were considered as potential OC candidates. 
We visually inspected the features of these star clusters, as described in 
Sect~\ref{confirm}. After this inspection, 704 OC candidates were 
proposed to be potentially real OCs.

Based on the same format as the table presented in previous 
works~\citep[e.g.,][]{castro2020,castro2021b}, we show the mean 
parameters and standard deviations 
(i.e., 1$\sigma$ errors) of each OC candidate in Table~\ref{Table:OCs}, 
including its right ascension (RA), declination (Dec), $l$, $b$, parallax 
($\varpi$), proper motions ($\mu_{\alpha^{*}}$ and $\mu_{\delta}$), 
age ($\log$($t$)), $RV$ ($V_{r}$), number of member stars ($N$), and its 
apparent angular size, which was computed by $\theta = 
\sqrt{\sigma_{l}^{2}+\sigma_{b}^{2}}$ following \cite{castro2020}. 
The full version of Table~\ref{Table:OCs} can be found online at the 
CDS~\footnote{\url{https://cdsarc.u-strasbg.fr/ftp/vizier.submit//new_oc_v3/}}, 
including the table of member stars of OCs.
For our newly found OCs, Fig.~\ref{fig:new_OCs} shows some 
examples of the distributions of the member stars in astrometric space 
and their CMDs. It is obvious that the OCs we proposed typically show a 
high concentration of member stars in all five astrometric parameters, 
especially the proper motions, and their member stars in the CMDs are 
almost concentrated on potentially theoretical isochrones.   
%


\subsection{Characteristics of the new OCs}
\label{sect:characteristics_oc}

As shown in Fig.~\ref{fig:parallax}, we made a parallactic comparison 
between the newly found OCs and the previously known OCs in
\cite{cantat2020b}, which shows a high level of consistency.
For the newly found OCs, the mean parallaxes of vast  majority (99.3$\%$) 
range from 0.06 to 3.00~mas, which yield distances from $\sim$ 330~pc 
to $\sim$ 16~kpc when estimating the distance as the inverse of the mean 
parallax. However, we note that this method is not ideal as it ignores the 
parallax uncertainties \citep[e.g.,][]{lindegren2021b}. In addition, $\sim$ 15\% 
of the OCs are closer than 1~kpc, $\sim$ 34\% are located between 1 and 
2~kpc, $\sim$ 37\% between 2 and 4~kpc, and $\sim$ 14\% of the OCs 
are further than 4~kpc. 
Among the newly found OCs, 214  of them have stars with \textit{RV}s in 
\textit{Gaia} EDR3, of which 71 OCs contain more than two stars with 
\textit{RV}s. 

For the newly found OCs, $\sim$ 85\% are located at Galactic latitudes 
of $|b| \le $ 5$^{\circ}$, $\sim$ 9\% are located in 5$^{\circ}$ < $|b|$ 
$\le$ 10$^{\circ}$ and only $\sim$ 6\% are located at Galactic latitudes 
$|b| > \ $10$^{\circ}$, showing they are concentrated in the Galactic disk. 
The distributions of the new OCs in the Galactic plane are shown in 
Fig.~\ref{fig:lb}, which indicates that the distribution of the new OCs follows 
a particularly similar distribution to the previously reported ones. 
Fig.~\ref{fig:age_r_z} presents the distribution of the new OCs and known 
OCs in the galactocentric radius ($R$) and altitude ($z$) above the Galactic 
middle plane, showing they are very consistent, and this result is also 
concordant with \cite{castro2021b}.
Hence, all of the above distributions of our newly found hidden OCs 
significantly densify the distribution constructed by previously known OCs.

The ages of the proposed OCs are color-coded in Fig.~\ref{fig:age_r_z}, 
which reveals the consistency of the age estimations with the previously 
known OC population compiled by \cite{cantat2020b}.
Besides, it can be seen that these new OCs gradually migrated further from 
the Galactic disk as they age, which has been mentioned by \cite{hao2021}.

The distribution of the 299 newly found OCs with ages younger than 20 
million years projected on to the Galactic plane is shown in 
Fig.~\ref{fig:young_ocs}, where the spiral arms defined by masers associated 
with high-mass star formation regions with Very Long Baseline Interferometry 
parallax measurements \citep{reid2019} are also displayed for comparison. 
More than 70\% of the young OCs are located in the spiral arms 
outlined by high-mass star formation region masers or bright O-B 
type stars \citep{xu2018,xu2021}, especially the more populous ones and the 
distribution of young OCs is similar to those of O-B type stars in \textit{Gaia}.
Hence, our result also confirmed that the young OCs are good tracers of 
the Galactic spiral structure, as described in some recent 
works~\citep[e.g.,][]{hao2021,castro2021a,poggio2021}.



\section{Conclusion}
\label{sect:conclusions}

Our proposed methodology, an amended version of sample-based clustering 
search method, which possess high spatial resolution, combined with the 
high-precision astrometric and photometric data in \textit{Gaia} EDR3, allowed 
us to find many previously unnoticed and hidden OCs in the 5D 
astrometric space ($l$, $b$, $\varpi$, $\mu_{\alpha^{*}}$, and $\mu_{\delta}$). 
In this work, the detection of over-density structures was performed in different 
distance bins as well as small spatial regions and multiple sample sizes were 
used for regions with different stellar densities, which do contribute to search 
for undetectable OCs in the Galactic disk. 
Then, the stellar clusters were reduced and confirmed as potentially real OCs 
based on their distributions, proper-motion dispersions, CMDs and 
\textit{RV}s, if available. 

Our blind search of the Galactic disk produced 1 930 known OCs and 82 known 
GCs, as well as 704 potentially new OCs. The new OCs have greatly improved 
the completeness of the Galactic OC census, and they commendably extend the
distribution of known OCs. Indeed, the high re-detection effectiveness for 
GCs also proves the reliability of our proposed method.
%

\section*{Acknowledgements}
We appreciate the anonymous referee for the instructive comments 
which help us to improve the paper.
This work was funded by NSFC Grants 11933011, 11873019, 11673066 and
11988101, and the Key Laboratory for Radio Astronomy. 
The authors thank Dr. Castro-Ginard for kindly 
providing the data of their newly found open clusters in~\citet{castro2021b}.
YJL thanks support from the Natural Science Foundation of Jiangsu Province 
(grant number BK20210999).
This work has made use of data from the European Space Agency (ESA) 
mission Gaia (\url{https://www.cosmos.esa.int/gaia}), processed by the Gaia 
Data Processing and Analysis Consortium 
(DPAC, \url{https://www. cosmos.esa.int/web/gaia/dpac/consortium}). 
Funding for the DPAC has been provided by national institutions, in particular 
the institutions participating in the Gaia Multilateral Agreement.


%

\begin{thebibliography}{1}
\expandafter\ifx\csname natexlab\endcsname\relax\def\natexlab#1{#1}\fi
\bibitem[Altman(1992)]{Altman1992} Altman, N. S. 1992, BAN, 46, 175
\bibitem[Andrae et al.(2018)]{Andrae2018} Andrae, R., Fouesneau, M., Creevey, O., et~al. 2018, \aap, 616, A8
\bibitem[Barnes(2007)]{Barnes2007} Barnes, Sydney A. 2007, \apj, 669, 1167
\bibitem[Baumgardt \& Hilker(2018)]{baumgardt2018} Baumgardt, H. \& Hilker, M. 2018, \mnras, 478, 1520
\bibitem[Baumgardt et al.(2019)]{baumgardt2019} Baumgardt, H., Hilker, M., Sollima, A. \& Bellini, A. 2019, \mnras, 482, 5138
\bibitem[Bertelli Motta et al.(2017)]{Motta2017} Bertelli, Motta C., Salaris, M., Pasquali, A. \& Grebel, E. K. 2017, \mnras, 466, 2161
\bibitem[Bonatto(2019)]{Bonatto2019} Bonatto, C. 2019, \mnras, 483, 2758
\bibitem[Bossini et al.(2019)]{Bossini2019} Bossini, D., Vallenari, A., Bragaglia, A., et al. 2019, \aap, 623, A108
\bibitem[Bressan et al.(2012)]{bressan2012} Bressan, A., Marigo, P., Girardi, L., et~al. 2012, \mnras, 427, 127
\bibitem[Buckner \& Froebrich(2014)]{buckner2014} Buckner, A. S.~M. \& Froebrich, D. 2014, \mnras, 444, 290
\bibitem[Cantat-Gaudin \& Anders(2020)]{cantat2020a} Cantat-Gaudin, T. \& Anders, F. 2020, \aap, 633, A99
\bibitem[Cantat-Gaudin et al.(2014)]{cantat2014} Cantat-Gaudin, T., Vallenari, A., Zaggia, S., et~al. 2014, \aap, 569, A17
\bibitem[Cantat-Gaudin et al.(2018)]{cantat2018} Cantat-Gaudin, T., Jordi, C., Vallenari, A., et~al. 2018, \aap, 618, A93
\bibitem[Cantat-Gaudin et al.(2019)]{cantat2019} Cantat-Gaudin, T., Krone-Martins, A., Sedaghat, N., et~al. 2019, \aap, 624, A126
\bibitem[Cantat-Gaudin et al.(2020)]{cantat2020b} Cantat-Gaudin, T., Anders, F., Castro-Ginard, A., et al. 2020, \aap, 640, A1
\bibitem[Cardelli et al.(1989)]{Cardelli1989} Cardelli, J.~A., Clayton, G.~C. \& Mathis, J.~S. 1989, \apj, 345, 245
\bibitem[Castro-Ginard et al.(2018)]{castro2018} Castro-Ginard, A., Jordi, C., Luri, X., et~al. 2018, \aap, 618, A59
\bibitem[Castro-Ginard et al.(2019)]{castro2019} Castro-Ginard, A., Jordi, C., Luri, X., Cantat-Gaudin, T. \& Balaguer-Nunez, L. 2019, \aap, 627, A35
\bibitem[Castro-Ginard et al.(2020)]{castro2020} Castro-Ginard, A., Jordi, C., Luri, X., et~al. 2020, \aap, 635, A45
\bibitem[Castro-Ginard et al.(2021a)]{castro2021a} Castro-Ginard, A., McMillan, P. J., Luri, X. et~al. 2021a, \aap, 652, A162
\bibitem[Castro-Ginard et al.(2021b)]{castro2021b} Castro-Ginard, A., Jordi, C., Luri, X., et~al. 2021b, arXiv: 2111.01819v1
\bibitem[Dias et al.(2002)]{dias2002} Dias, W.~S., Alessi, B.~S., Moitinho, A. \& L{\'e}pine, J. R.~D. 2002, \aap, 389, 871
\bibitem[Dias et al.(2021)]{dias2021} Dias, W. S., Monteiro, H., Moitinho, A., et al. 2021, \mnras, 504, 356
\bibitem[Donati et al.(2014)]{donati2014} Donati, P., Cantat-Gaudin, T., Bragaglia, A., et~al. 2014, \aap, 561, A94
\bibitem[Ester et al.(1996)]{ester1996} Ester, M., Kriegel, H.-P., Sander, J. \& Xu X. 2020, 2nd International Conference on Knowledge Discovery and Data Mining (AAAI Press), 96, 226
\bibitem[Evans et al.(2018)]{evans2018} Evans, D.~W., Riello, M., De Angeli, F., et~al. 2018, \aap, 616, A4
\bibitem[Ferreira et al.(2019)]{ferreira2019} Ferreira, A.~F., Santos, J.~F.~C., Corradi, W.~J.~B., Maia, F.~F.~S. \& Angelo, M.~S.  2019, \mnras, 483, 5508
\bibitem[Ferreira et al.(2020)]{ferreira2020} Ferreira, A.~F., Corradi, W.~J.~B., Maia, F.~F.~S., Angelo, M.~S. \& Santos, J.~F.~C. 2020, \mnras, 496, 2021
\bibitem[Ferreira et al.(2021)]{ferreira2021} Ferreira, A.~F., Corradi, W.~J.~B., Maia, F.~F.~S., Angelo, M.~S. \& Santos, J.~F.~C. 2021, \mnras, 502, L90
\bibitem[Friel(1995)]{friel1995} Friel, E.~D. 1995, \araa, 33, 381-414
\bibitem[\textit{Gaia} Collaboration et al.(2016)]{Prusti2016} \textit{Gaia} Collaboration (Prusti, T., et al.)\ 2016, \aap, 595, A1
\bibitem[\textit{Gaia} Collaboration et al.(2018)]{Brown2018} \textit{Gaia} Collaboration (Brown, A.~G.~A., et al.)\ 2018, \aap, 616, A1
\bibitem[\textit{Gaia} Collaboration et al.(2020)]{gaia2020} \textit{Gaia} Collaboration (Brown, A.~G.~A., et al.)\ 2020, \aap, 649, A1
\bibitem[Hao et al.(2020)]{hao2020} Hao, C.~J., Xu, Y., Wu, Z.~Y., et~al. 2020, \pasp, 132, 034502
\bibitem[Hao et al.(2021)]{hao2021} Hao, C. J., Xu, Y., Hou, L. G., et~al. 2021, \aap, 652, A102
\bibitem[Hatzidimitriou et al.(2019)]{hatzidimitriou2019} Hatzidimitriou, D., et~al. 2019, \aap, 626, A90
\bibitem[He et al.(2021)]{he2021} He, Z.~H., Xu, Y., Hao, C.~J., Wu, Z.~Y. \& Li, J.~J. 2021, Res. Astron. Astrophys. 21, 093
\bibitem[Hunt \& Reffert(2021)]{hunt2021} Hunt E. L. \& Reffert S. 2021, \aap, 646, A104
\bibitem[Janes \& Adler(1982)]{janes1982} Janes, K. \& Adler, D. 1982, \apjs, 49, 425
\bibitem[Kharchenko et al.(2013)]{kharchenko2013} Kharchenko, N.~V., Piskunov, A.~E., Schilbach, E., R{\"o}ser, S. \& Scholz, R.-D. 2013, \aap, 558, A53
\bibitem[Lampe et al.(2011)]{Lampe2011} Lampe, O. D. \& Hauser, H. 2011, 2011 IEEE Pacific Visualization Symp. (Piscataway, NJ: IEEE), 171
\bibitem[Lapenna et al.(2015)]{lapenna2015} Lapenna, E., Origlia, L., Mucciarelli, A., et al. 2015, \apj, 798, 23
\bibitem[Lindegren et al.(2021a)]{lindegren2021a} Lindegren, L., Klioner, S.~A., Hern{\'a}ndez, J. et~al. 2021a, \aap, 649, A2
\bibitem[Lindegren et al.(2021b)]{lindegren2021b} Lindegren, L., Bastian, U., Biermann, M., et~al. 2021b, \aap, 649, A4
\bibitem[Liu \& Pang(2019)]{liu2019} Liu, L. \& Pang, X. 2019, \apjs, 245, 32
\bibitem[Marino et al.(2018)]{marino2018} Marino, A. F., Milone, A. P., Casagrande, L., et al. 2018, \apjl, 863, L33
\bibitem[Mermilliod et al.(2009)]{mermilliod2009} Mermilliod, J. C., Mayor, M. \& Udry, S. 2009, \aap, 498, 949
\bibitem[O'Donnell(1994)]{ODonnell1994} O'Donnell, J.~E. 1994, \apj, 422, 158
\bibitem[Overbeek et al.(2017)]{overbeek2017} Overbeek, J.~C., Friel, E. D., Donati, P., et~al. 2017, \aap, 598, A68
\bibitem[Perren et al.(2015)]{Perren2015} Perren, G. I., V{\'a}zquez, R. A. \& Piatti, A. E. 2015, \aap, 576, A6
\bibitem[Poggio et al.(2021)]{poggio2021} Poggio, E., Drimmel, R., Cantat-Gaudin, T., et al. 2021, \aap, 651, A104
\bibitem[R{\"o}ser et al.(2010)]{roser2010} R{\"o}ser, S., Demleitner, M. \& Schilbach, E. 2010, \aj, 139, 2440
\bibitem[Reid et al.(2019)]{reid2019} Reid, M.~J., Menten, K.~M., Brunthaler, A., et al.  2019, \apj, 885, 131
\bibitem[Sim et al.(2019)]{sim2019} Sim, G., Lee, S. H., Ann, H. B. \& Kim, S. 2019, J. Korean Astron. Soc., 52, 145
\bibitem[Skrutskie et al.(2006)]{skrutskie2006} Skrutskie, M.~F., Cutri, R. M., Stiening, R., et~al. 2006, \aj, 131, 1163
\bibitem[Soubiran et al.(2018)]{soubiran2018} Soubiran, C., Cantat-Gaudin, T., Romero-G{\'o}mez, M., et~al. 2018, \aap, 619, A155
\bibitem[Tarricq et al.(2021)]{tarricq2021} Tarricq, Y., Soubiran, C., Casamiquela, L. et al. 2021, \aap, 647, A19
\bibitem[Vandenberg(1983)]{vandenberg1983} Vandenberg, E.~D. 1983, \apjs, 51, 29
\bibitem[Vereshchagin \& Chupina(2016)]{vereshchagin2016} Vereshchagin, S.~V. \& Chupina, N.~V. 2016, Baltic Astron., 25, 432
\bibitem[Xu et al.(2018)]{xu2018} Xu, Y., Bian, S. B., Reid, M. J., et al. 2018, \aap, 616, L15
\bibitem[Xu et al.(2021)]{xu2021} Xu, Y., Hou, L. G., Bian, S. B., et al. 2021, \aap, 645, L8

\end{thebibliography}
%

\end{document}